\documentclass[aps,twocolumn]{revtex4}
\usepackage[utf8]{inputenc}
 \usepackage{amsmath,amssymb}
 \usepackage{graphicx}
 \usepackage{float}
\usepackage{color}
\usepackage[colorlinks=true,citecolor=blue,linkcolor=blue,urlcolor=black]{hyperref}    
\begin{document}

\title{Big-Bang Nucleosynthesis and WIMP Dark Matter Freeze-Out as Probes of Yukawa
Cosmology}

\author{Ava Shahbazi Sooraki\footnote{awashahbazi@hafez.shirazu.ac.ir} and Ahmad Sheykhi
\footnote{asheykhi@shirazu.ac.ir}}

\address{Department of Physics, College of
Science, Shiraz University, Shiraz 71454, Iran\\
Biruni Observatory, College of Science, Shiraz University, Shiraz
71454, Iran}

\begin{abstract}
We investigate Big-Bang Nucleosynthesis (BBN) in the context of
Yukawa cosmology. We first derive the modified Friedmann equations
by starting from the first law of thermodynamics on the apparent
horizon. Using observational data on \(^4\mathrm{He}\), deuterium,
and \(^7\mathrm{Li}\) abundances, we place stringent bounds on the
Yukawa coupling \(\alpha\). The \(^4\mathrm{He}\) and deuterium
constraints are mutually consistent (\(-0.24 \lesssim \alpha
\lesssim 0.12\)), while \(^7\mathrm{Li}\) requires $\alpha \in
[-0.76,\,-0.72]$. This indicate that Yukawa cosmology cannot
resolve the Lithium Problem. We then extend our analysis to WIMP
freeze-out, and show that the modified Hubble parameter alters the
relic abundance, yielding an independent constraint \(-0.017
\lesssim \alpha \lesssim 0.018\) from \(\Omega_{\mathrm{CDM}}h^2 =
0.120 \pm 0.001\). We also derive the modified time-temperature
relation, and show that the positive \(\alpha\) raises the early
Universe temperature. Our analysis demonstrates that BBN and dark
matter relic abundance serve as complementary probes of modified
gravity. Our studies confirm that Yukawa cosmology is a testable
framework for early-Universe physics.
\end{abstract}
\maketitle
\newpage
\section{Introduction\label{Intro}}
The current standard model of cosmology, the Lambda-Cold Dark
Matter ($\Lambda$CDM) model, has been remarkably successful in
describing a vast range of phenomena, from the large-scale
structure of the cosmic web to the temperature anisotropies of the
cosmic microwave background (CMB) radiation \cite{Sakr2022,
Akarsu2023}. At its core, $\Lambda$CDM posits the existence of two
mysterious components: cold dark matter (CDM), which interacts
only gravitationally and drives structure formation, and the
cosmological constant ($\Lambda$), a form of dark energy
responsible for the observed accelerated expansion of the
universe. Despite its empirical successes, the model faces
profound theoretical challenges \cite{Sakr2022, Akarsu2023,
Turner2026,Blanchard2025}. The most notable of these is the
cosmological constant problem. This is the staggering discrepancy
of some $120$ orders of magnitude between the observed value of
$\Lambda$ and the vacuum energy density predicted by quantum field
theory. This discrepancy strongly suggests that our understanding
of gravity on cosmological scales may be incomplete. This tension
has motivated a rich field of research into modified theories of
gravity, which seek to explain cosmic acceleration without
invoking a fundamental cosmological constant. Among the numerous
proposed alternatives, those introducing an additional, or
modified, gravitational interaction in the form of a Yukawa
potential have garnered significant attention
\cite{Koyama2016,Clifton2012, Carroll2005}. In such "Yukawa
cosmology" models, the standard Newtonian potential, $\Phi_{\rm
Newton}=-GM/r$, is augmented by a corrective term
\cite{Nishonov2026,DeMartino2018, AraujoFilho2024, DelValle2018}
\begin{equation*}
\Phi(r)=-\frac{GM}{r}\left(1+\alpha e^{-r/\lambda}\right),
\end{equation*}
where $\alpha$ represents the coupling strength of the new
interaction relative to gravity, and $\lambda$ is its
characteristic range (or Compton wavelength). Thus, a point
particle with mass $m$ at distance $r=R$ from mass $M$, feels the
force $ \vec{F}=-m \vec{\nabla} \phi(r)|_{r=R}$. This modification
can arise from various theoretical frameworks, including
scalar-tensor theories, $f(R)$ gravity, and theories with large
extra dimensions, where a massive scalar degree of freedom
mediates an additional force. Based on the modified Newton's law
of gravitation, and applying the notion of the entropic force
scenario proposed by Verlinde \cite{Ver} and developed in
\cite{Cai2,sheyECFE}, one can derive the corrections to the
entropy associated with the horizon of spacetime \cite{KS,
Sheykhi:2025}. These correction terms alter the dynamical field
equations of gravity through the thermodynamics-gravity
conjecture. The key observation is that any modification to the
gravitational force law implies a corresponding modification to
the entropy associated with the cosmological horizon. Starting
from a generalized entropy expression, one can apply the first law
of thermodynamics, $dE = T dS + W dV$, to the apparent horizon of
a Friedmann-Robertson-Walker universe. Following this procedure
the standard Friedmann equations acquire correction terms that
depend explicitly on the parameters characterizing the entropy
modification \cite{sheyECFE, wang1, Cai22}. On cosmological
scales, such a modification alters the expansion history of the
universe and the growth rate of density perturbations, providing a
potential dynamical explanation for the observed cosmic
acceleration. While Yukawa-type corrections are often
parameterized and tested on solar system and galactic scales,
their implications for the very early universe remain a powerful
and, as yet, less explored probe. The earliest cosmological epoch
for which we have precise observational data is the period of Big
Bang Nucleosynthesis (BBN). Occurring in the first few minutes
after the Big Bang, BBN was the process by which protons and
neutrons fused into the first light atomic nuclei: deuterium $D$,
$_{}^{3}\textit{He}$, $_{}^{4}\textit{He}$, and
$_{}^{7}\textit{Li}$ \cite{RH1, RH2, Planck, Brain}. The
primordial abundances of these elements are extremely sensitive to
the universal expansion rate at that time, which is governed by
the Friedmann equation . A modified gravitational theory, such as
Yukawa cosmology, will inevitably alter this expansion rate by
modifying the effective gravitational constant $G_{\rm eff}$ or by
introducing new, dynamical degrees of freedom that contribute to
the total energy density of the universe \cite{Anish,Luciv,
Sahoo2, Ava1}. Furthermore, this framework allows us to revisit
the persistent cosmological lithium problem, i.e., the discrepancy
between the predicted primordial abundance of $_{}^{7}{\rm Li}$
and its observed value in metal-poor halo stars \cite{Coc}. By
examining whether the modified expansion rate induced by the
Yukawa potential can bring the theoretical lithium abundance into
agreement with observations, we assess whether Yukawa cosmology
offers a viable resolution to this long-standing puzzle.
Consequently, the precisely measured primordial abundances act as
a formidable and independent constraint on any physics beyond the
standard models of particle physics and cosmology. Any deviation
from the standard BBN prediction, which is based on General
Relativity and the known particle content of the Standard Model,
would signal new physics. Conversely, the remarkable agreement
between the observationally inferred abundances and the standard
BBN predictions allows us to place stringent limits on any such
deviations. On the other hand Dark matter (DM) remains one of the
biggest unsolved puzzles in modern physics, eluding detection in
direct and indirect detection experiments while collider searches
have yet to identify a statistically significant signal. For
recent reviews on the topic of DM, see
Refs.~\cite{Bertone:2004pz,Feng:2010gw,Gelmini:2015zpa,Roszkowski:2017nbc,Arcadi:2017kky,Battaglieri:2017aum,
Munoz:2003gx, Taoso:2007qk, Jungman:1995df}. One robust
measurement we do have is the DM relic density, $\Omega_{\rm
CDM}h^2 = 0.120 \pm 0.001$~\cite{Planck}, which has been a key
constraint in model building. Among the most compelling DM
candidates are weakly interacting massive particles (WIMPs). Their
thermal relic abundance naturally matches the observed DM density
for weak-scale masses and annihilation cross sections, the
so-called WIMP miracle. Many extensions of the Standard Model
accommodate WIMP candidates, making it essential to understand how
modifications to the standard cosmological picture affect
freeze-out predictions \cite{Mod90}. In the standard cosmology,
freeze-out occurs during the radiation-dominated era, and the
relic abundance is determined by the annihilation cross section
and the expansion rate of the Universe~\cite{Kolb:1988aj}.

However, because we lack direct observations prior to Big Bang
Nucleosynthesis (BBN), the expansion history before BBN is not
firmly established. Alternative cosmological models including
modified gravity~\cite{Catena:2004ba,Catena:2006bd,Kang},
quintessence~\cite{Salati:2002md,Profumo:2003hq}, brane-world
scenarios~\cite{Okada:2004nc}, and early matter-dominated
epochs~\cite{Bernal:2018ins} can alter the Hubble expansion rate
and thus affect the relic abundance. The thermal relic abundance
of WIMPs is particularly sensitive to the cosmological expansion
rate during freeze-out. Since the freeze-out condition is governed
by the competition between the annihilation rate of dark matter
particles and the Hubble expansion rate, any departure from the
standard cosmological evolution prior to BBN modifies the
freeze-out dynamics and consequently the predicted relic
abundance~\cite{Kolb:1988aj}. Therefore, the observed dark matter
relic density provides a powerful and independent cosmological
probe of modified gravity scenarios and non-standard expansion
histories~\cite{Catena:2004ba,Catena:2006bd,Kang,Salati:2002md,Profumo:2003hq,Okada:2004nc,Bernal:2018ins}.
In this paper, we derive the modified Friedmann equations from the
first law of thermodynamics, which differs from the entropic force
approach used in \cite{Jusufi2023}. A different result is obtained
relative to theirs, featuring a logarithmic correction term.
Building upon this thermodynamic framework we perform a detailed
analysis to constrain the parameters of Yukawa cosmology using the
latest observations of primordial element abundances. We will
compute the modified expansion history during the BBN epoch within
the context of a Yukawa gravitational potential, derive the
resulting primordial light-element yields, and compare them with
observational data. This comparison will allow us to delineate the
allowed regions in the $\alpha$-$\lambda$ parameter space. By
leveraging the precision of BBN, this work provides a unique and
powerful test of modified gravity theories in the high-redshift,
high-temperature universe, complementing and extending constraints
obtained from late-time cosmological probes and local laboratory
experiments. We also examine the lithium problem within the
framework of Yukawa cosmology.
Furthermore we investigate the thermal freeze-out of WIMPs within
the Yukawa cosmology framework. The modified expansion history
predicted by this framework is incorporated into the Boltzmann
equation governing the evolution of the WIMP number density. We
derive an analytical expression for the relic abundance as a
function of the Yukawa parameter $\alpha$ and obtain an
independent constraint on $\alpha$ by requiring consistency with
the observed dark matter relic density. These constraints provide
a complementary test of Yukawa cosmology alongside those derived
from Big Bang Nucleosynthesis. Additionally, we will examine the
relationship between cosmic time and temperature in the early
universe, in the framework of Yukawa cosmology. The present work
therefore provides a unified analysis of Yukawa cosmology by
confronting the modified expansion history with two independent
cosmological observables, namely primordial nucleosynthesis
and the dark matter relic abundance.\\
This paper is organized as follows. In Section II, we derive the
modified entropy from the Yukawa potential using Verlinde's
entropic force scenario and obtain the modified Friedmann
equations via the first law of thermodynamics. Section III applies
these modifications to Big Bang Nucleosynthesis, deriving
constraints on the Yukawa parameters from the primordial
abundances of deuterium, $_{}^{4}\textit{He}$, and
$_{}^{7}\textit{Li}$. In Section IV, we investigate the thermal
freeze-out of weakly interacting massive particles within the
Yukawa cosmology framework, derive the corresponding relic
abundance, and obtain constraints on the Yukawa parameter from the
observed dark matter relic density. Section V examines the
modified time-temperature relation. Finally, we summarize our
results and closing remarks in Section VI.
\section{Entropic Corrections to Friedmann Equations via Yukawa Potential}
\label{Sect:II}
In this section, we derive the modified Friedmann
equations for Yukawa cosmology. We begin by obtaining the entropy
correction induced by the Yukawa modification to Newton's law,
following Verlinde's entropic force scenario. We then apply the
first law of thermodynamics on the apparent horizon of a FRW
universe to obtain the corrected Friedmann equations.
\subsection{Modified Entropy from Yukawa Potential}
Following Verlinde's entropic force scenario, when a test particle
moves away from the holographic screen, the entropic force on this
body has the form \cite{Ver}
\begin{equation}
F\Delta x = T \Delta S.
\end{equation}
where $\Delta x$ is the displacement of the particle from the
holographic screen, and $T$ and $\Delta S$ are the temperature and
entropy change on the screen, respectively. For Einstein's
gravity, the entropy-area relation is $S = A/4$, where $A = 4\pi
R^2$ is the area of the horizon. To incorporate modifications
arising from the Yukawa potential, we consider the generalized
entropy expression \cite{sheyECFE, Sheykhi:2021}
\begin{equation}
    S_h = \frac{A}{4G} + \mathcal{S}(A),
\end{equation}
When the second term is absent, the above equation reduces to the
standard Bekenstein-Hawking entropy. The discrete spectrum of the
horizon area implies that the entropy changes in units of $\Delta
S$, with
\begin{equation}
    dS_h = \frac{\partial S}{\partial A} dA = \left[\frac{1}{4G} + \frac{\partial \mathcal{S}}{\partial A}\right] dA.
\end{equation}

The energy of the surface $\Sigma$ is identified with the
relativistic rest mass $E = M$. According to Verlinde's
holographic picture, the area $A$ is related to the number of bits
$N$ on the screen by $A = QN$, where $Q$ is a fundamental
constant. Using the equipartition law of energy, the total energy
on the surface is \cite{Padmanabhan:2004}
\begin{equation}
E = \frac{1}{2} N k_B T.
\end{equation}
The entropic force follows from Eq.~(1), where $\Delta S$
corresponds to one fundamental unit of entropy when the
displacement is $|\Delta x| = \eta \lambda_m$, with the entropy
gradient directed radially inward from the surface. Taking $\Delta
N = 1$ and $\Delta A = Q$, we get
\begin{equation}\label{F3}
    F = -\frac{GMm}{R^2}\left(\frac{Q}{2\pi k_B \eta}\right)\left[\frac{1}{4G} + \frac{\partial \mathcal{S}}{\partial A}\right]_{A = 4\pi R^2}.
\end{equation}
This expression reduces to Newton's law of gravitation to first
order provided we set $\eta = 1/8\pi k_B$ and $Q = 1$. We thus
obtain
\begin{equation}\label{F4}
F = -\frac{GMm}{R^2}\left[\frac{1}{G} + 4\,\frac{\partial \mathcal{S}}{\partial A}\right]_{A = 4\pi R^2}.
\end{equation}

Let us take the following  Yukawa-type gravitational potential
\begin{equation}
\phi(r) = -\frac{GMm}{r}\left(1 + \alpha
e^{-r/\lambda}\right)\bigg|_{r=R},
\end{equation}
where \(\alpha\) is the dimensionless Yukawa coupling parameter.
Note that $ \alpha > 0$ corresponds to an additional attractive
force beyond Newtonian gravity, while  $-1 < \alpha < 0$
introduces a repulsive correction that weakens the gravitational
interaction. The standard Newtonian limit is recovered for
\(\alpha = 0\). The wavelength of massive graviton reads  $\lambda
= \hbar/(m_g c) > 10^{20}$ m, which leads to $m_g < 10^{-64}$ kg
for the graviton mass \cite{Visser:1998}. Such a potential arises
from the Einstein field equations $G_{\mu\nu} = 8\pi
G(T^{\text{matter}}_{\mu\nu} + T_{\mu\nu}^{\text{field}})$ by
linearizing the geometry $g_{\mu\nu} = \eta_{\mu\nu} + h_{\mu\nu}$
and assuming a non-zero graviton mass. In the weak-field limit $(h
\ll 1)$, this yields a Yukawa-like potential $\phi(r) \sim
(\text{const}/r) e^{-r/\lambda}$ \cite{Visser:1998}. The full
potential around a mass $M$ is then the superposition of the
standard Newtonian term and the Yukawa correction, with
$GMm\alpha\sim\text{const}$. We note that such Yukawa
modifications also appear in modified gravity theories such as
$f(R)$ gravity \cite{Capozziello:2009, Berezhiani2009,
Benisty2022}.

Applying $F = -\nabla \phi(r)|_{r=R}$ to the Yukawa potential yields the modified Newtonian gravitational force
\begin{equation}
F = -\frac{GMm}{R^2}\left[1 + \alpha\left(\frac{R +
\lambda}{\lambda}\right)e^{-R/\lambda}\right].
\end{equation}
In the entropic force scenario, we can see that the ratio of the modified to
Newtonian force equals the ratio of the modified to
Bekenstein-Hawking entropy gradient. Comparing the above equation with the standard Newtonian force $F_N = -GMm/R^2$, we define the correction factor

\begin{equation}\label{corr}
    \mathcal{F}(R) \equiv 1 + \alpha e^{-R/\lambda}\left(1 + \frac{R}{\lambda}\right).
\end{equation}
Consequently, the entropy gradient associated with the apparent horizon is modified as
\begin{equation}\label{dS}
    \frac{dS_h}{dR} = \frac{2\pi R}{G}\, \mathcal{F}(R)=\frac{2\pi
        R}{G}\left[1 + \alpha e^{-R/\lambda}\left(1 +
    \frac{R}{\lambda}\right)\right].
\end{equation}
Integrating Eq.~\ref{dS}, we obtain the modified form of the
entropy associated with the apparent horizon, which takes the form
\begin{equation}\label{Sh}
    S_h = \frac{\pi R^2}{G} - \frac{2\pi \alpha}{G} e^{-R/\lambda}\left(R^2+3\lambda R + 3\lambda^2\right).
\end{equation}
Where we set the integration constant $S_0 = \pi R^2/G$ so that
$S_h \to \pi R^2/G$ as $R \to \infty$ (since $e^{-R/\lambda} \to
0$), recovering the area law in the limit $\alpha \to 0$. It is
evident that $\lambda$ carries the dimension of length, while
$\alpha$ is dimensionless. Exponential corrections to the entropy
can be seen in \cite{Chatterjee:2020}. Consequently, the parameter
$\alpha$ introduces modifications to the Newtonian dynamics. This
analysis demonstrates that an entropy modification of the form
given in Eq.~\eqref{Sh} necessarily implies a corresponding
modification to Newton's law of gravitation. This establishes an
effective modified gravity framework. The additional entropy term
can be interpreted as a volume-law entanglement contribution
arising from graviton degrees of freedom. Within this
interpretation, the Yukawa coupling $\alpha$ originates from the
entanglement entropy associated with the volume-law scaling.
\subsection{Derivation of Modified Friedmann Equations through Yukawa Cosmology }
We now derive the modified Friedmann equations arising from the
Yukawa correction to the gravitational force. Following
\cite{Sheykhi:2025}, we apply the thermodynamics-gravity
conjecture to the apparent horizon of a FRW universe via the first
law of thermodynamics. The Yukawa modification to the Newtonian
potential alters the entropy-area relation of cosmological
horizons (Eq. \eqref{Sh}), which in turn modifies the
gravitational field equations through the holographic principle.
Note that the area of the apparent horizon is $A = 4\pi R^2$, and
the volume enveloped by the apparent horizon is $V = 4\pi R^3/3$.
The apparent horizon radius $R$ is defined as \cite{Hay1,Hay2,Bak}
\begin{equation}\label{radius}
R = \frac{1}{\sqrt{H^2 + k/a^2}},
\end{equation}
where $a(t)$ is the scale factor, $H = \dot{a}/a$ is the Hubble
parameter, and $k = -1,0,1$ corresponds to open, flat, and closed
universes, respectively. The background spacetime is described by
the FRW metric
\begin{equation}
ds^2 = -dt^2 + \frac{a^2(t)}{1 - kr^2} dr^2 + a^2(t) r^2 (d\theta^2 + \sin^2\theta d\phi^2).
\end{equation}
The temperature associated with the apparent horizon is \cite{Hay1,Hay2,Bak}
\begin{equation}\label{Th}
T_h = \frac{\kappa}{2\pi} = -\frac{1}{2\pi R}\left(1 - \frac{\dot{R}}{2HR}\right),
\end{equation}
where $\kappa$ is the surface gravity. For $\dot{R} < 2HR$, one
may take $T = |\kappa|/2\pi$ to avoid negative temperature. The
matter content of the universe is assumed to be a perfect fluid
with energy-momentum tensor $T_{\mu\nu} = (\rho + p)u_\mu u_\nu +
p g_{\mu\nu}$, where $\rho$ and $p$ are the energy density and
pressure, respectively. The conservation of energy in the FRW
background gives the continuity equation
\begin{equation}\label{cont}
\dot{\rho} + 3H(\rho + p) = 0.
\end{equation}
Due to the expansion of the universe, a work term appears in the
first law of thermodynamics. The corresponding work density for a
perfect fluid is \cite{Hay2}
\begin{equation}\label{work}
W = \frac{1}{2}(\rho - p).
\end{equation}
The total energy inside the apparent horizon is $E = \rho V = \frac{4\pi}{3}R^3\rho$, whose differential is
\begin{equation}\label{dE}
dE = 4\pi R^2 \rho \, dR + \frac{4\pi}{3}R^3 d\rho.
\end{equation}
Using the continuity equation (\ref{cont}) to express $d\rho = -3H(\rho + p) dt$, we obtain
\begin{equation}\label{dE2}
dE = 4\pi R^2 \rho \, dR - 4\pi H R^3 (\rho + p) dt.
\end{equation}
Now we apply the first law of thermodynamics on the apparent
horizon, $dE = T_h dS_h + W dV$. Substituting Eqs.~(\ref{dS}),
(\ref{Th}), (\ref{work}), and (\ref{dE2}) into the first law,
multiplying by $G$ and rearranging terms yields
\begin{align}\label{eq2}
4\pi G H R^3 (\rho + p) dt = \left(1 - \frac{\dot{R}}{2HR}\right) \mathcal{F}(R) dR \\
+ 2\pi G R^2 (\rho + p) dR.
\end{align}
Now we employ the continuity equation (\ref{cont}) to replace
$\rho + p = -\dot{\rho}/(3H)$, and substituting into
Eq.~(\ref{eq2}) while using $dR = \dot{R} dt$ and dividing by $dt$
gives
\begin{equation}\label{eq5}
-4\pi G R^3 \dot{\rho} + \frac{2\pi G}{H} R^2 \dot{\rho} \dot{R} = 3\left(1 - \frac{\dot{R}}{2HR}\right) \mathcal{F}(R) \dot{R}.
\end{equation}
Factoring $-4\pi G R^3 \dot{\rho}$ on the left yields
\begin{equation}\label{eq7}
-4\pi G R^3 \dot{\rho} \left(1 - \frac{\dot{R}}{2HR}\right) = 3\left(1 - \frac{\dot{R}}{2HR}\right) \mathcal{F}(R) \dot{R}.
\end{equation}
Assuming $\left(1 - \frac{\dot{R}}{2HR}\right) \neq 0$ (which
holds in an expanding universe), we can cancel this common factor,
leading to the simplified relation
\begin{equation}\label{keyeq}
-\frac{2}{R^3} \mathcal{F}(R) dR = \frac{8\pi G}{3} d\rho.
\end{equation}
This equation directly links the entropy correction to the energy
density without any further assumptions. To make progress, we
expand $\mathcal{F}(R)$ for the early universe where $R/\lambda
\ll 1$. Using the expansion $e^{-R/\lambda} = 1 - R/\lambda +
R^2/(2\lambda^2) - \cdots$, we find
\begin{align}\label{fexpand}
\mathcal{F}(R)= 1 + \alpha\left(1 - \frac{R^2}{2\lambda^2} + \cdots\right)=(1 + \alpha) \\
- \frac{\alpha}{2}\frac{R^2}{\lambda^2} \notag
+ \mathcal{O}(R^3/\lambda^3)+... .
\end{align}
Neglecting terms of order $1/\lambda^3$ and higher, and substituting into Eq.~(\ref{keyeq}) we get
\begin{equation}\label{eq11}
-\frac{2(1+\alpha)}{R^3} dR + \frac{\alpha}{\lambda^2 R} dR = \frac{8\pi G}{3} d\rho.
\end{equation}
Integrating both sides yields
\begin{equation}\label{eq12}
\frac{1+\alpha}{R^2} + \frac{\alpha}{\lambda^2} \ln R=\frac{8\pi G}{3} \rho  + \frac{\Lambda}{3},
\end{equation}
where $\Lambda$ is an integration constant which can be
interpreted as the cosmological constant. Now, using the
definition of the apparent horizon radius , we have $    R = (H^2
+ k/a^2)^{-1/2}$, so $\ln R = -\ln (H^2+k/a^2)/2$. Substituting
into above equation gives
\begin{equation}\label{final}
\left(H^2+\frac{k}{a^2}\right)-\Gamma \ln
\left(H^2+\frac{k}{a^2}\right) = \frac{8\pi
G_{\text{eff}}}{3}(\rho+\rho_{\Lambda}) .
\end{equation}
Where we have defined $G_{\text{eff}}=G/{(1+\alpha)}$,
$\rho_{\Lambda}=\Lambda/(8\pi G)$ and
$\Gamma=\frac{\alpha}{2\lambda^2 (1+\alpha)}$. Equation
(\ref{final}) is the modified first Friedmann equation inspired by
the Yukawa potential. When $\alpha = 0$, it reduces to the
standard Friedmann equation. The second Friedmann equation can be
derived by combining Eq.~(\ref{final}) with the continuity
equation (\ref{cont}). Differentiating with respect to time and
using $\dot{H} = dH/dt$, for a flat universe ($k=0$) we obtain
\begin{equation}\label{accel}
\dot{H} \left(1-\frac{\alpha}{2\lambda^2 (1+\alpha)}
H^{-2}\right)= -\frac{4\pi G}{1+\alpha} (\rho + p).
\end{equation}
When $\alpha = 0$, the second Friedmann equation reduces to its
standard general relativity (GR) form. In the early universe,
where $H$ is large, the logarithmic term in Eq.~(\ref{final})
becomes negligible relative to $H^2$, so we obtain approximately
\begin{equation}\label{highen}
H^2 \approx \frac{8\pi G}{3(1+\alpha)} \rho.
\end{equation}
For a radiation-dominated universe ($\omega = 1/3$), the
continuity equation gives $\rho = \rho_0 a^{-4}$, leading to the
scale factor evolution $a(t) \sim t^{1/2}$ (since $H = \dot{a}/a
\sim t^{-1/2}$). This is the same power law as in standard
cosmology, but with an effective Newton constant $G_{\text{eff}} =
G/(1+\alpha)$. For $\alpha > 0$, the expansion rate is reduced
compared to standard cosmology. Thus, from the first law of
thermodynamics on the apparent horizon, we obtain the modified
Friedmann equations for Yukawa cosmology. The results depend on
$\alpha$ and $\lambda$ (or equivalently $m_g = \hbar/(\lambda
c)$), and reduce to standard general relativity for $\alpha = 0$.
\section{Big Bang nucleosynthesis constraints on Yukawa cosmology\label{BBN}}
\subsection{Modified Expansion Rate and Amplification Factor}
In this section, we investigate the consequences of Yukawa
cosmology during the radiation-dominated epoch and analyze its
effects on Big Bang Nucleosynthesis. Our approach is to find a
perturbative solution of the modified Friedmann equation by
treating the logarithmic correction term as a small perturbation.
This enables us to derive the corrections to the Hubble expansion
rate and to define an effective amplification factor that
quantifies the deviation from standard cosmology. We begin with
the modified Friedmann equation obtained from the Yukawa-corrected
entropy in Eq. \eqref{final}. For a spatially flat universe
($k=0$) and neglecting the cosmological constant
($\rho_{\Lambda}=0$) during the early stages of cosmic evolution,
Eq. \eqref{final} simplifies to
\begin{equation} \label{Fr5}
H^2-\Gamma \ln H^2=\frac{H_{GR}^2}{1+\alpha},
\end{equation}
where $H_{GR}^2=8\pi G\rho/3$. Since the correction term $\Gamma
\ln H^2$ is small compared to $H^2$, we seek an approximate
solution to first order in $\Gamma$. To facilitate this, we
rewrite Eq. \eqref{Fr5} as
\begin{equation} \label{Fr}
H^2-\epsilon=C \quad\Longrightarrow\quad H=(C+\epsilon)^{1/2},
\end{equation}
with $C\equiv H_{GR}^2/(1+\alpha)$. where $C=H_{GR}^2/(1+\alpha)$
and $\epsilon=\Gamma \ln H^2$.  Inserting $H^2 = C + \epsilon$
into the Eq. \eqref{Fr5} gives
\begin{equation} \label{Ff1}
\epsilon-\Gamma \ln(C+\epsilon)=0.
\end{equation}
Since $\epsilon \ll C$, we can approximate the logarithm by its
Taylor expansion to find
\begin{equation} \label{Ff2}
\ln(C+\epsilon)\simeq \ln C+\frac{\epsilon}{C}+\mathcal{O}\left(
\frac{\epsilon^2}{C^2} \right).
\end{equation}
Inserting the above equation into Eq. \eqref{Ff1} we arrive at
\begin{multline} \label{Fs}
\epsilon-\Gamma \left(\ln C+\frac{\epsilon}{C} \right)=0, \; \;
\Rightarrow  \; \; \epsilon \left(1-\frac{\Gamma}{C}
\right)=\Gamma \ln C.
\end{multline}
Solving for $\epsilon$ we obtain
\begin{equation} \label{Ffa}
\epsilon=\frac{\Gamma \ln C}{1-\Gamma/C}.
\end{equation}
Since $\Gamma/C \ll 1$, we expand the denominator to linear order,
obtaining
\begin{equation} \label{Ffs}
\epsilon\simeq \Gamma \ln C\left(1+ \frac{\Gamma}{C} \right)\simeq
\Gamma \ln
 C+\mathcal{O}(\Gamma^2).
\end{equation}
Substituting this result into Eq. \eqref{Fr} and keeping terms to
first order in $\Gamma$, we obtain $H^2\simeq C(1+({\Gamma}/{C}
)\ln C)$. After expanding and restoring the definitions of $C$ and
$\Gamma$, the modified Hubble parameter takes the form
\begin{equation} \label{Ffd}
H\simeq \frac{H_{GR}}{\sqrt{1+\alpha}}\left[1+ \Gamma \;
\frac{1+\alpha}{2H_{GR}^2}
\ln\left(\frac{H_{GR}^2}{1+\alpha}\right) \right],
\end{equation}
For the leading-order solution, the modified Hubble parameter
satisfies $H<H_{\rm GR}$ for $\alpha>0$, while $H>H_{\rm GR}$ for
$-1<\alpha<0$, under the assumption that the logarithmic
correction remains perturbatively small, i.e. $\Gamma/H_{\rm
GR}^{2}\ll1$. It is convenient to express the modified Hubble
parameter in terms of the standard GR result by introducing an
amplification factor $Z(T)$
\begin{equation}
H(T) \equiv Z(T) H_{GR}(T).
\end{equation}
Comparing this with Eq. \eqref{Ffd}, we identify
\begin{equation} \label{ZT1}
Z(T)=\frac{1}{\sqrt{1+\alpha}}\left[1+ \Gamma \;
\frac{1+\alpha}{2H_{GR}^2}
\ln\left(\frac{H_{GR}^2}{1+\alpha}\right) \right]
\end{equation}
To study BBN in Yukawa cosmology, we require the temperature
dependence of the amplification factor. Inserting the relativistic
particle energy density $\rho(T) = \pi^2 g_* T^4/30$, with $g_*
\simeq 10$, the standard Hubble parameter $H_{GR}^2 = 8\pi
G\rho/3$ and $\Gamma$ into Eq. \eqref{ZT1}, we arrive at
\begin{equation}\label{zt}
Z(T)=\frac{1}{\sqrt{1+\alpha}}\left[1+ \frac{45\alpha}{16\lambda^2
\pi^3GgT^4} \; \ln\left(\frac{4\pi^3GgT^4}{45(1+\alpha)}\right)
\right]
\end{equation}
The above expression represents the amplification factor in Yukawa
cosmology, which is dimensionless in natural units ($\hbar = c =
k_B = 1$). As expected, in the limit $\alpha \to 0$, we have
$\Gamma \to 0$ and hence $Z(T) \to 1$, recovering the standard GR
expansion rate.
\subsection{ Observational Bounds from {\textit{ $_{}^{4}\textrm{He}$, D} and \textit {Li} } Abundances in Yukawa Cosmology}
We now proceed to constrain the Yukawa parameter $\alpha$ using
BBN observables. The abundances of ${}^2\textit{H}$,
${}^4\textit{He}$, and  $_{}^{7}\textit{Li}$ are sensitive probes
of the early-universe expansion rate, making them ideal for
testing the modified cosmology developed above. By comparing the
theoretical abundance predictions with precision observational
data, we derive bounds on $\alpha$. The key novelty of our
analysis lies in replacing the conventional $Z$-factor which in
standard cosmology parametrizes the effective number of neutrino
species \cite{Luciano}, with the modified amplification factor
$Z(T)$ from Eq. \eqref{zt}. In standard cosmology, $Z$ is related
to $N_{\nu}$ by \cite{Anish,Luciano,Boran}
\begin{align}
    Z_{\nu} = \left[ 1 + \frac{7}{43}(N_{\nu} - 3) \right]^{1/2},
\end{align}
with $Z=1$ corresponding to $N_{\nu}=3$. The baryon-to-photon
ratio $\eta_{10} \equiv 10^{10} n_B/n_\gamma$ \cite{Adv,Simha}
serves as an additional input parameter. To isolate the effects of
Yukawa cosmology, we fix $N_{\nu}=3$ throughout our analysis. This
ensures that any deviation of $Z$ from unity is uniquely
attributed to the modified gravitational dynamics encoded in
$Z(T)$, rather than to exotic particle content. The primordial
abundances produced during BBN are sensitive to the baryon density
and the expansion rate of the early Universe. The former is
quantified by the baryon-to-photon ratio, which compares the
number density of baryons to that of CMB photons. The coupled
nonlinear differential equations governing BBN do not admit
analytical solutions. Consequently, relating the theoretical
predictions for the light-element abundances to the input
parameters, namely the expansion rate $Z$ and the baryon-to-photon
ratio $\eta_B$, requires numerical computations. However, by
mapping the predicted abundance contours for ${}^4\textit{He}$,
$D$, and ${}^7\textit{Li}$ as functions of $Z$ and $\eta_B$, one
finds that the abundances vary smoothly with these parameters,
allowing for simple fits to the numerical results. Although the
dependence on $\eta_B$ is not strictly linear, linear fits in both
$Z$ and $\eta_B$ provide adequate approximations for our purposes.
We adopt the methodology of Refs. \cite{Anish, Sahoo}, which we
briefly summarize below.

\paragraph{Helium-4 abundance.} The abundance of ${}^4\textit{He}$.
The primordial ${}^4\textit{He}$ mass fraction is given by the numerical best-fit formula \cite{Kneller,Annu}
\begin{equation} \label{bestfit}
Y_p = 0.2485 \pm 0.0006 + 0.0016 \left[ (\eta_{10} - 6) + 100 (Z -
1) \right],
\end{equation}
where $\eta_{10} \equiv 10^{10} \eta_B$ is the baryon density
parameter, with $\eta_B \equiv n_B/n_\gamma\simeq 6$
\cite{Adv,Simha,Wamp}. The amplification factor $Z$ is given by
Eq. \eqref{zt}, encoding the modified expansion dynamics of Yukawa
cosmology. In the standard GR limit $Z=1$, Eq. \eqref{bestfit}
yields $(Y_p)|_{\text{GR}} = 0.2485 \pm 0.0006$. Observations of
primordial ${}^4\text{He}$, assuming $\eta_{10}=6$, give $Y_p =
0.2449 \pm 0.004$ \cite{Brain}. Equating this with Eq.
\eqref{bestfit} yields
\begin{equation}
0.2449 \pm 0.0040 = 0.2485 \pm 0.0006 + 0.0016 \left[ 100(Z - 1) \right],
\end{equation}
which implies
\begin{equation} \label{zhe}
Z = 1.0475 \pm 0.105.
\end{equation}

\paragraph{Deuterium abundance.} Primordial deuterium is synthesized via $n+p \rightarrow{}^2\text{H}+\gamma$.
The numerical fit of Ref.~\cite{Adv} gives
\begin{equation} \label{zde}
Y_{D_p} = 2.6(1 \pm 0.06) \left(\frac{6}{\eta_{10} - 6(Z - 1)}\right)^{1.6}.
\end{equation}
Setting $Z=1$ and $\eta_{10}=6$ recovers the GR prediction
$Y_{D_p}|_{\text{GR}} = 2.6 \pm 0.16$. Matching this with the
observed value $Y_{D_p} = 2.55 \pm 0.03$ \cite{Brain} yields
\begin{equation}
Z = 1.062 \pm 0.444, \label{zobs}
\end{equation}
which exhibits partial agreement with the helium-4 constraint from
Eq.~\eqref{zhe}.

\paragraph{Lithium-7 abundance.} Although the baryon density parameter $\eta_{10}$
successfully reconciles the predicted and observed abundances of
${}^4\textit{He}$ and $D$, it fails to do so for ${}^7\textit{Li}$.
Standard BBN calculations with $\eta_{10} \simeq 6$ accurately
reproduce the observed yields of deuterium and helium-4, but
systematically predict a ${}^7\textit{Li}$ abundance that is
larger than observations by a factor of 3-4. This discrepancy,
quantified in Ref.~\cite{Boran} as
\begin{align*}
\frac{\text{Li}|_{\text{GR}}}{\text{Li}|_{\text{obs}}} \in [2.4, 4.3],
\end{align*}
constitutes the well-known \textit{Cosmological Lithium Problem}.
Remarkably, while the standard BBN framework is extraordinarily
successful in predicting the primordial abundances of
${}^4\textit{He}$ and $D$, it fails to account for the observed
${}^7\textit{Li}$ abundance in metal-poor halo stars. This
unresolved tension presents a significant challenge to our
understanding of early-universe physics \cite{Boran}. The
numerical fit for the primordial ${}^7\textit{Li}$ abundance is
given by
\begin{equation}\label{dd}
Y_{\text{Li}} = 4.82(1 \pm 0.1)\left[\frac{\eta_{10} - 3(Z - 1)}{6}\right]^2.
\end{equation}
Comparing this theoretical prediction with the observational value
$Y_{\text{Li}} = 1.6 \pm 0.3$ \cite{Brain} yields the constraint
\begin{equation} \label{lit}
Z = 1.960025 \pm 0.076675.
\end{equation}
This result is incompatible with the bounds derived from
${}^4\textit{He}$ and $D$ in Eqs. \eqref{zhe} and\eqref{zobs},
highlighting the Lithium problem in standard cosmology. We will
analyze the implications of this result for Yukawa cosmology in
the following subsection.
\subsection{Confronting Yukawa Cosmology with BBN Observations and the Lithium Problem}
\begin{figure}[H]
\includegraphics[scale=0.88]{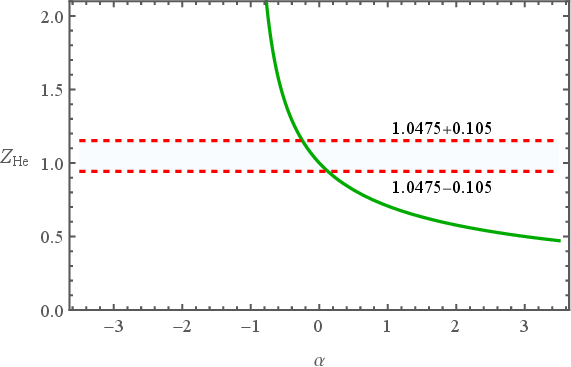}
\caption{$Z_{He}$ vs Yukawa parameter $\alpha $. The observational
interval from Eq.~\eqref{zhe} is shown as a horizontal band. We
have set $\eta_{10}=6$ and the freeze-out temperature $T_f=1
MeV$.} \label{Fig1}
\end{figure}

\begin{figure}[H]
\includegraphics[scale=0.86]{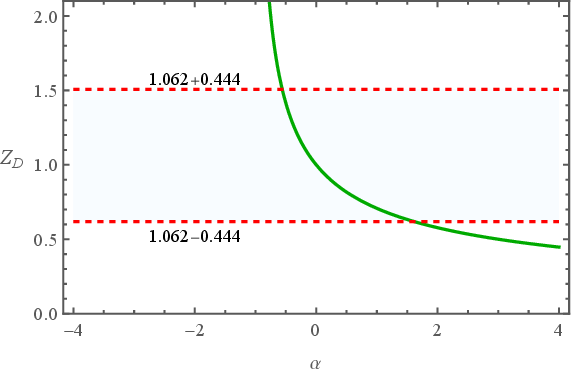}
\caption{$Z_D$ vs Yukawa parameter $\alpha$. The observational
interval from Eq.~\eqref{zobs} is shown as a horizontal band. We
have set $\eta_{10}=6$ and the freeze-out temperature $T_f=1
MeV$.}
 \label{Fig2}
\end{figure}

We now summarize the main results of BBN analysis and derive the
corresponding constraints on the Yukawa parameter $\alpha$. Fig
\ref{Fig1} shows the relation between the amplification factor $Z$
and $\alpha$ as given by Eq. \eqref{zt}, together with the
observational band from Eq. \eqref{zhe}. The intersection of these
curves directly translates into the following bound on $\alpha$
from ${}^4\textit{He}$ measurements
\begin{equation} \label{dhes}
    -0.24 \lesssim \alpha \lesssim 0.12.
\end{equation}
Where we have taken $M_p = (8\pi G)^{-1/2} \simeq 2.4 \times 10^{18}$ GeV for the numerical evaluation \cite{Luciv}.
Applying the same procedure to deuterium, we use the observational
constraint on $Z$ from Eq.~\eqref{zobs} together with the
theoretical relation $Z(\alpha)$ from Eq.~\eqref{zt}.
Fig~\ref{Fig2} shows the resulting comparison, from which we
obtain the following bound on $\alpha$
\begin{equation} \label{deuts}
-0.5 \lesssim \alpha \lesssim 1.5.
\end{equation}
For values of $\alpha$ within the range given by
Eq.~\eqref{deuts}, the Yukawa cosmology predicts a deuterium
abundance that is consistent with observations. Since this
interval overlaps with the tighter constraint from
$_{}^{4}\textit{He}$ (Eq.~\eqref{dhes}), their intersection,
$-0.24 \lesssim \alpha \lesssim 0.12,$ defines the region of
parameter space where the model simultaneously reproduces the
observed primordial abundances of both deuterium and
$_{}^{4}\textit{He}$. This $\alpha$ range is also consistent with
other bounds in the literature \cite{Gonzalez2023, AboHasan2023,
Jusufi2023}.

\begin{figure}[H]
\includegraphics[scale=0.86]{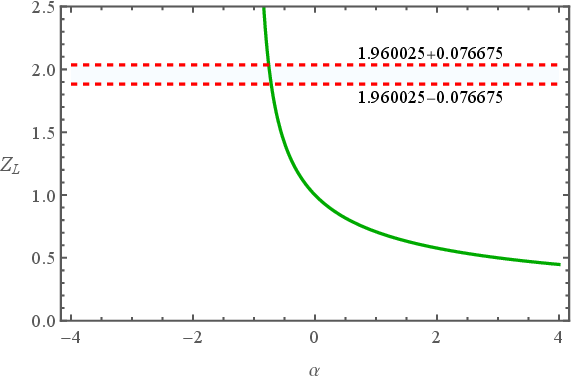}
\caption{$Z_{Li}$ vs Yukawa parameter $\alpha $. The observational
interval from Eq.~\eqref{lit} is shown as a horizontal band. We
have set $\eta_{10}=6$ and the freeze-out temperature $T_f=1
MeV$.} \label{Fig3}
\end{figure}

Turning now to lithium, we present in Fig.~\ref{Fig3} the
constraint on $\alpha$ obtained from the primordial
${}^7\textit{Li}$ abundance. Following the same procedure as for
$_{}^{4}\textit{He}$ and deuterium, we combine the observational
bound on $Z$ from Eq.~\eqref{lit} with the theoretical relation
$Z(\alpha)$ from Eq.~\eqref{zt}. This yields
\begin{equation} \label{ZLi}
 -0.76 \lesssim \alpha \lesssim -0.72.
\end{equation}
Fig~\ref{Fig3} clearly demonstrates that the parameter space
permitted by ${}^7\textit{Li}$ observations is completely disjoint
from the regions allowed by ${}^4\textit{He}$ (Eq.~\ref{dhes}) and
deuterium (Eq.~\ref{deuts}). The cosmological lithium problem
arises from the discrepancy between the primordial
${}^7\textit{Li}$ abundance predicted by standard BBN and the
observed value. If a modified cosmological model could
simultaneously account for the ${}^4\textit{He}$ and $D$ abundance
which are already well described by standard BBN, and the observed
lithium abundance, this would offer a resolution to the problem
without invoking astrophysical mechanisms. Our analysis shows that
while the Yukawa parameter $\alpha$ can successfully accommodate
both  ${}^4\textit{He}$ and deuterium observations, no single
value of $\alpha$ can simultaneously reproduce the measured
lithium abundance. Consequently, there is no overlap between the
$\alpha$ ranges favored by ${}^4\textit{He}$/$D$ and that required
by ${}^7\textit{Li}$, meaning the lithium discrepancy persists
within this framework. This indicates that further modifications
beyond the present Yukawa framework may be required to address the
lithium problem. It should be emphasized that the numerical ranges
in Eqs.~\eqref{dhes}, \eqref{deuts}, and \eqref{ZLi} are obtained
for the specific choice $\lambda \sim 10^{26}\,\text{m}$ and $T =
1\,\text{MeV}$ and may shift at different energy scales. However,
the qualitative conclusion remains robust i.e. the lithium-favored
$\alpha$ interval shows no overlap with those from
${}^4\textit{He}$ and deuterium across the relevant BBN
temperature range, and the deviations from $\Lambda$CDM
predictions remain minimal.
\section{Dark Matter Relic Abundance as a Probe of Yukawa Gravity} \label{sec:DM}
Recent cosmological observations have established a precise
measurement of the cold dark matter abundance ~\cite{Planck}
\begin{equation}\label{OmegaObs}
0.119 \le \Omega_{\rm dm} h^2 \le 0.121,
\end{equation}
which provides a sensitive probe of any modification to the
expansion history of the Universe. In this section, we investigate
the thermal relic abundance of weakly interacting massive
particles (WIMPs) within the Yukawa cosmology introduced in
Sec.~II. We derive an independent constraint on the Yukawa
parameter $\alpha$ by incorporating the modified expansion rate
into the Boltzmann equation governing thermal freeze-out. The
resulting relic abundance is then compared with the observed dark
matter density to determine the allowed range of $\alpha$.

We assume that dark matter consists of a stable particle species
$\chi$ with mass $m_\chi$ and thermally averaged annihilation
cross section $\langle\sigma v\rangle$. We consider a generic
weakly interacting massive particle (WIMP) as the dark matter
candidate, assuming its mass lies in the range
$m_\chi=(100$--$500)\,\mathrm{GeV}$ and that the typical
annihilation cross section is of the order of
$\sigma\sim\alpha_{\rm em}^{2}/M_{\rm ew}^{2}$, where $\alpha_{\rm
em}=1/137$ is the electromagnetic fine-structure constant and
$M_{\rm ew}\sim100\,\mathrm{GeV}$ denotes the electroweak scale.
The evolution of the WIMP number density is governed by the
Boltzmann equation~\cite{Kolb:1988aj,Feng:2003zu}
\begin{equation}\label{Boltzmann}
\dot n_\chi + 3H n_\chi=-\langle \sigma v \rangle
\left(n_\chi^2 - n_{\chi,{\rm EQ}}^2\right),
\end{equation}
where $n_\chi$ is the number density of dark matter particles,
$n_{\chi,{\rm EQ}}$ denotes the equilibrium number density, and
$H$ is the Hubble expansion rate. The thermal average of the total
annihilation cross section times the relative velocity $\langle
\sigma v \rangle$ is given by
\begin{equation}
\langle \sigma v \rangle=\frac{1}{n_{EQ}^2}\int\frac{d^3p_1}{(2\pi)^3}\frac{d^3p_2}{(2\pi)^3}f(E_1) f(E_2) \sigma v,
\label{sigma_avg}
\end{equation}
where $f(E)$ is the fermion distribution function,
$f(E)=1/(1+\exp(E/T))$. Finally the number density at equilibrium
is given by
\begin{equation}\label{nEQ}
n_{EQ}=\int\frac{d^3p}{(2\pi)^3}f(E).
\end{equation}
Following the standard treatment of thermal relics
\cite{Kolb:1988aj}, we introduce the dimensionless variables
\begin{equation}\label{xY}
x \equiv \frac{m_\chi}{T},
\qquad
Y \equiv \frac{n_\chi}{s},
\end{equation}
where $T$ is the temperature and $s$ is the entropy density,
\begin{equation}
s=\frac{2\pi^2}{45}h_*T^3,\label{entropy}
\end{equation}
with $h_*$ being the number of relativistic degrees of freedom for
entropy density. Assuming entropy conservation and using
Eq.~(\ref{xY}), the Boltzmann equation becomes
\begin{equation}\label{BoltzmannY}
\frac{dY}{dx}=-\frac{s}{xH}\langle \sigma v\rangle\left(Y^2 - Y_{\rm EQ}^2\right).
\end{equation}
For nonrelativistic particles ($x\gg3$), the equilibrium abundance
is well approximated by~\cite{Kolb:1988aj}
\begin{equation}
Y_{\rm EQ}\simeq g\frac{45}{2\pi^4}\left(\frac{\pi}{8}\right)^{1/2}\frac{x^{3/2}e^{-x}}{h_*},
\label{Yeq}
\end{equation}
where $g=2$ is the spin polarizations of the dark matter particle.
The annihilation cross section can be parameterized as
\begin{equation}
\langle \sigma v \rangle=\sigma_0 x^{-l},
\label{sigmav}
\end{equation}
where $l=0$ corresponds to $s$-wave annihilation and $l=1$
corresponds to $p$-wave annihilation. Substituting
Eq.~(\ref{sigmav}) into Eq.~(\ref{BoltzmannY}), the Boltzmann
equation takes the final compact form
\begin{equation}
\frac{dY}{dx}=-\lambda x^{-l-2}\left(Y^2 - Y_{\rm EQ}^2\right),
\label{BoltzmannFinal_GR}
\end{equation}
where $\lambda$ is a constant given by
\begin{equation}
\lambda=\left(\frac{x\langle\sigma v\rangle s}{H(m)}\right)_{x=1}=0.264\left(\frac{h_*}{\sqrt{g_*}}\right)M_{\rm Pl}m_\chi\sigma_0.
\label{lambda}
\end{equation}
\subsection{Effects of Modified Expansion on WIMP Freeze-Out}
In the Yukawa cosmology developed in Sec.~II, the Hubble parameter
during the radiation-dominated epoch, after neglecting the
logarithmic correction term in the modified Friedmann equation, is
given by
\begin{equation}
H(T)\simeq\frac{1}{\sqrt{1+\alpha}}H_{\rm GR}(T),
\label{HY1}
\end{equation}
In standard cosmology during the radiation dominated era, the Hubble
parameter as a function of the temperature is given by
$H_{GR}(T)=1.67g_*^{1/2}T^{2}/M_{p}$ ~\cite{Kolb:1988aj}, so the above equation can be rewritten as
\begin{equation}
H(T)\simeq\frac{1.66\,g_*^{1/2}}{\sqrt{1+\alpha}}\frac{T^2}{M_{\rm Pl}}.
\label{HYT}
\end{equation}
Equation~(\ref{HYT}) shows that the Yukawa modification changes
only the overall normalization of the Hubble expansion rate, while
preserving the standard temperature dependence during the
radiation-dominated era,
\begin{equation}
H(T)\propto\frac{1}{\sqrt{1+\alpha}}T^2.
\label{HT2}
\end{equation}
Expressing the temperature in terms of the variable $x=m_\chi/T$, one obtains
\begin{equation}
H(x)\simeq\frac{H_{GR}(m_\chi)}{\sqrt{1+\alpha}}x^{-2},
\label{HYx}
\end{equation}
where
\begin{equation}
H_{GR}(m_\chi)=1.66\,g_*^{1/2}\frac{m_\chi^2}{M_{\rm Pl}}.
\end{equation}
Substituting Eqs.~(\ref{entropy}), (\ref{sigmav}) and (\ref{HYx}) into Eq.~(\ref{BoltzmannY}), we obtain
\begin{equation}
\frac{dY}{dx}=-\tilde\lambda x^{-\tilde{l}-2}\left(Y^2 - Y_{\rm EQ}^2\right),
\label{BoltzmannFinal}
\end{equation}
where $\tilde\lambda$ is given by
\begin{multline}
\tilde\lambda= \left(\frac{x\langle\sigma v\rangle
s}{H_{Mod}(m)}\right)_{x=1}=\frac{H_{GR}(m)}{H_{Mod}(m)}\lambda\\=
0.264 \left(\frac{h_*}{\sqrt{g_*}}\right)M_{\rm Pl}m_\chi
\tilde\sigma_0. \label{lambdatilde}
\end{multline}
Here we have introduced the effective quantities $\tilde{l}$ and
$\tilde{\sigma}_0$ in order to express the modified Boltzmann
equation in the same form as in standard cosmology. Since the
Yukawa modification rescales the Hubble expansion rate only by the
constant factor $H_{\rm Mod}=H_{\rm GR}/\sqrt{1+\alpha}$, the
standard temperature dependence $H\propto T^2$ during the
radiation-dominated era remains unchanged. Using $T=m_\chi/x$,
together with $s\propto T^3\propto x^{-3}$ and $\langle\sigma
v\rangle=\sigma_0x^{-l}$, one finds $s\langle\sigma
v\rangle/(xH)\propto x^{-l-2}$, exactly as in the standard
cosmology. Therefore, the Yukawa correction modifies only the
overall normalization of the Boltzmann equation while leaving its
$x$-dependence unchanged. Consequently, the effective annihilation
exponent remains identical to that of the standard cosmology,
$\tilde{l}=l$, and the entire effect of the modified expansion
history can be absorbed into the redefinition of the effective
annihilation cross section
\begin{equation}
\tilde\sigma_0=
\frac{H_{GR}(m)}{H_{Mod}(m)} \sigma_0=\sqrt{1+\alpha}\,\sigma_0,
\label{sigma0tilde}
\end{equation}
Therefore, within the approximation where the logarithmic
correction term in the modified Friedmann equation is neglected,
the entire gravitational effect of Yukawa cosmology is encoded in
the redefinition of the annihilation cross section,
$\sigma_0 \to \tilde\sigma_0 = \sqrt{1+\alpha}\,\sigma_0$. The
temperature dependence of the freeze-out process, characterized by
the exponent $l$, remains unchanged compared to standard cosmology
($\tilde l = l$).
\subsection{Freeze-out and Relic Density in Modified Yukawa Cosmology}
We can obtain an approximate analytical solution of the Boltzmann
equation by the following arguments. Initially, for large
temperatures the annihilation rate is larger than the expansion
rate of the universe and the WIMP abundance follows the
equilibrium abundance. At some point $x_f$ the annihilation rate
becomes comparable to the expansion rate and the dark matter
particle decouples from the thermal bath. Freeze-out occurs when
the interaction rate becomes comparable to the expansion rate
$\Gamma    =n_{\chi,{\rm EQ}}  \langle \sigma v\rangle \simeq  H$
~\cite{Kolb:1988aj}. After freeze out for $x \gg x_f$ we can
neglect the $Y_{EQ}$ term in the Boltzmann equation, and
Eq.~(\ref{BoltzmannFinal}) reduces to ${dY}/{dx}=  -\tilde\lambda
x^{-l-2}Y^2.$ Then the equation can be easily integrated and the
solution $Y_{\infty}\equiv Y(x=\infty)$ is given by
\begin{equation}
Y_{\infty}=\frac{l+1}{\tilde{\lambda}}x_f^{\,l+1},
\end{equation}
where the freeze-out temperature $x_{f}$ is given by \cite{Kolb:1988aj}
\begin{equation}
x_f=\ln A-\left(l+\frac12\right)\ln(\ln A),\label{xf}
\end{equation}
where
\begin{equation}
A=0.038(l+1)\left(\frac{g}{\sqrt{g_*}}\right)M_{\rm Pl}m_\chi \tilde{\sigma}_0.
\label{A}
\end{equation}
For $\alpha=0$, Eq.~(\ref{A}) reduces to the well-known expression
of the standard dark matter freeze-out formalism in general
relativity~\cite{Kolb:1988aj, Kang}. After having integrated the
Boltzmann equation for $Y(x)$, then the relic abundance for the
dark matter particle is given by
\begin{equation}
\Omega_{dm}h^{2}=\frac{mY_{\infty}s(T_{0})h^{2}}{\rho_{cr}},
\end{equation}
where $T_{0}$ is the today's temperature. Here we make use of the following values
\begin{align*}
T_{0} & = 2.73~{\rm K} = 2.35\times10^{-13}~{\rm GeV},\\
h_{*}(T_{0}) & = 3.91,\\
\rho_{cr}/h^{2} & = 8.1\times10^{-47}~{\rm GeV}^{4}.
\end{align*}
The present relic density in the standard cosmology is therefore~\cite{Kolb:1988aj, Kang}
\begin{equation}
\Omega_\chi h^2=1.07\times10^9\,\frac{(l+1)x_f^{\,l+1}}{(h_*/\sqrt{g_*})M_p\sigma_0}\,{\rm GeV}^{-1},
\label{OmegaGR}
\end{equation}
Substituting Yukawa modified quantities into the standard relic
density formula, i.e. replacing $\sigma_0 \to \tilde\sigma_0$ and
$\lambda \to \tilde\lambda$ in Eq.~(\ref{OmegaGR}), we obtain the
relic density in Yukawa cosmology
\begin{equation}
\Omega_\chi h^2(\alpha)=1.07\times10^9\,\frac{(l+1)x_f^{\,l+1}}{(h_*/\sqrt{g_*})M_p \tilde\sigma_0}\,{\rm GeV}^{-1}.
\label{OmegaGeneral}
\end{equation}
Using Eq.~(\ref{sigma0tilde}), one finally obtains the relic density in Yukawa cosmology
\begin{equation}
\Omega_\chi h^2(\alpha)=1.07\times10^9\,\frac{(l+1)x_f^{\,l+1}}{(h_*/\sqrt{g_*})M_p\sqrt{1+\alpha} \,\sigma_0}\,{\rm GeV}^{-1}.
\label{OmegaYukawa}
\end{equation}
Equation (\ref{OmegaYukawa}) is the principal result of this
section. It shows that the Yukawa modification enters the relic
abundance solely through the rescaling of the annihilation cross
section, $\sigma_0 \to \sqrt{1+\alpha}\,\sigma_0$, while leaving
the temperature dependence of the Boltzmann equation unchanged.
Thus, for given $l$, $\sigma_0$, and $m_\chi$, the relic density
is obtained as an analytical function of $\alpha$ from
Eq.~(\ref{OmegaYukawa}). Since the temperature dependence of the
Boltzmann equation remains unchanged, the observed dark matter
abundance can be used to place direct constraints on the Yukawa
parameter $\alpha$ for fixed values of the WIMP mass and
annihilation cross section. For a benchmark WIMP mass of
\(m_\chi=100\,\mathrm{GeV}\), thermal freeze-out typically occurs
at $T_f \simeq {m_\chi}/{x_f} \sim 4-5~\mathrm{GeV},$
corresponding to \(x_f\simeq20-25\) . Following the standard
treatment of thermal relic freeze-out, we adopt the values $g_* =
h_* = 85$ for the numerical calculations ~\cite{Dodelson, Husdal,
Laine}. This choice is consistent with the Standard Model
thermodynamic degrees of freedom in the GeV-scale freeze-out
regime.
\subsection{Observational Bounds on the Yukawa Parameter from WIMP Relic Abundance}
\begin{figure}[H]
        \includegraphics[scale=0.85]{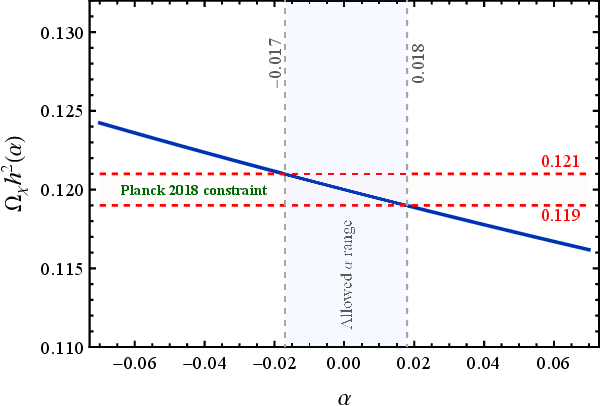}
        \caption{Dark matter relic abundance $\Omega_{\mathcal{X}}h^2$ versus $\alpha$. The horizontal band shows the observational constraint from Eq.~(\ref{OmegaObs}) with $g_*=h_*=85$ ($T_f\simeq5\,\mathrm{GeV}$).}
        \label{Fig4}
    \end{figure}

To obtain a direct bound on the Yukawa parameter, we assume a
benchmark thermal WIMP with \cite{Dodelson,Profumo, Smirnov,
Steigman, Drees}
\begin{equation}
m_\chi = 100~{\rm GeV},\qquad\langle \sigma v \rangle\simeq(2-  3)\times10^{-26}~{\rm cm^3\,s^{-1}},
\label{benchmark}
\end{equation}
which corresponds to $\sigma_0 =(1.7- 2.5) \times 10^{-9}~{\rm
GeV}^{-2}$ and a freeze-out parameter $ x_f\simeq20-25$ for
$s$-wave annihilation ($l=0$). To derive an approximate constraint
on the Yukawa parameter, we first neglect the weak logarithmic
dependence of the freeze-out parameter on the Yukawa correction
(see \eqref{xf}). We treat the freeze-out parameter $x_f$ as
approximately constant and approximate $x_f \simeq x_f^{(0)}$. In
this approximation, the relic density in Yukawa cosmology is given
by
\begin{equation}
\Omega_\chi h^2(\alpha)\simeq \frac{\Omega_\chi h^2(0)}{\sqrt{1+\alpha}},
\label{OmegaApprox}
\end{equation}
where $\Omega_\chi h^2(0)$ denotes the relic abundance in the
standard cosmology ($\alpha=0$). This approximation is justified
because $x_f$ depends only logarithmically on $\alpha$ through
$\tilde{\sigma}_0$ in Eq.~(\ref{A}), whereas the explicit
$\sqrt{1+\alpha}$ factor in the denominator dominates the
$\alpha$-dependence of the relic density for small $\alpha$.
Throughout this work, for the benchmark WIMP we normalize the
relic abundance in the standard cosmology to the observed value
$\Omega_\chi h^2(0) = 0.12$ \cite{Smirnov, Leane}. Using the
observational constraint from Eq.~(\ref{OmegaObs}),  we obtain
\begin{equation}
0.119 \le \frac{0.12}{\sqrt{1+\alpha}} \le 0.121.
\end{equation}
Solving for $\alpha$ gives
\begin{equation}
-0.15 \lesssim \alpha \lesssim 0.19.
\label{AlphaConstraint}
\end{equation}
This result should be compared with the BBN bounds obtained in
Sec.~III. For ${}^{4}{He}$ \eqref{dhes} and deuterium
\eqref{deuts}, which shows complete overlap with the WIMP
constraint. The intersection of these two bounds yields the final
allowed range
\begin{equation}
-0.15 \lesssim \alpha \lesssim 0.12.
\label{FinalAlpha}
\end{equation}
We therefore conclude that Yukawa cosmology with a standard
thermal WIMP is fully compatible with both BBN and dark matter
observations, provided that the Yukawa parameter lies in the range
given by Eq.~(\ref{FinalAlpha}). \\ To assess the impact of the
approximation $x_f\simeq x_f^{(0)}$, we retain the logarithmic
dependence of the freeze-out parameter on the Yukawa correction.
Substituting the Yukawa-modified quantity
$\tilde{A}=\sqrt{1+\alpha}\,A_0$ into Eq.~(\ref{xf}), we obtain
\begin{multline}
x_f(\alpha)=\ln\!\left(\sqrt{1+\alpha}\,A_0\right)\\
-\left(l+\frac12\right)\ln\! \left[\ln\!\left(\sqrt{1+\alpha}\,A_0\right) \right].
\label{xfExactAlpha}
\end{multline}
Using $\ln\!\left(\sqrt{1+\alpha}\,A_0\right)=\ln
A_0+(1/2)\ln(1+\alpha),$ Eq.~(\ref{xfExactAlpha}) becomes
\begin{multline}
x_f(\alpha)=\ln A_0+\frac12\ln(1+\alpha)\\
-\left(l+\frac12\right)\ln\!\left[\ln A_0+\frac12\ln(1+\alpha)\right].
\end{multline}
Since
$\left|{\ln(1+\alpha)}/{2\ln A_0}\right|\ll1, $ the last logarithm can be expanded according to
$\ln(a+\delta)=\ln a+{\delta}/{a}+\mathcal{O}(\delta^2),$ with $a=\ln A_0$ and $\delta=\ln(1+\alpha)/2$, Therefore
\begin{equation}
\ln\!\left[\ln A_0+\frac12\ln(1+\alpha)\right]\simeq\ln(\ln A_0)+\frac{\ln(1+\alpha)}{2\ln A_0}.
\end{equation}
Substituting this result back into Eq.~(\ref{xfExactAlpha}) yields
\begin{align}
x_f(\alpha)\simeq\;&
\ln A_0-\left(l+\frac12\right)\ln(\ln A_0)\nonumber\\
&+\frac12\left[1-\frac{l+\frac12}{\ln A_0}\right]
\ln(1+\alpha).
\end{align}
Using the definition of $x_f^{(0)}=\ln A_0-\left(l+1/2\right)\ln(\ln A_0)$, the above expression reduces to
\begin{equation}
x_f(\alpha)\simeq x_f^{(0)}+\frac12\left[1-\frac{l+\frac12}{\ln A_0}\right]\ln(1+\alpha).
\label{xfImproved}
\end{equation}
Since $\ln A_0\simeq x_f^{(0)}$ and, for the benchmark WIMP
considered here, $x_f^{(0)}\simeq 20-25$, the second term inside
the square brackets constitutes only a few-percent correction.
Therefore, for the $s$-wave annihilation case ($l=0$), the
freeze-out parameter is approximated by
\begin{equation}
x_f(\alpha)\simeq x_f^{(0)}+\frac12\ln(1+\alpha).
\label{xfApprox}
\end{equation}
Substituting Eq.~(\ref{xfApprox}) into the relic abundance formula, Eq.~(\ref{OmegaYukawa}), yields
\begin{equation}\label{OmegaImproved}
        \Omega_\chi h^2(\alpha)
        \simeq
        \frac{\Omega_\chi h^2(0)}{\sqrt{1+\alpha}}
        \left[1+\frac{1}{2x_f^{(0)}}\ln(1+\alpha)\right] .
\end{equation}
 Imposing the observational
constraint in \eqref {OmegaObs}, the allowed values of the Yukawa
parameter are determined numerically from
Eq.~(\ref{OmegaImproved}) by plotting $\Omega_\chi h^2(\alpha)$ as
a function of $\alpha$ (See Fig. \ref{Fig4}). The resulting
constraint is
    \begin{equation}
        -0.017
        \lesssim
        \alpha
        \lesssim
        0.018.
        \label{AlphaConstraintExact}
    \end{equation}
The bound in Eq.~(\ref{AlphaConstraintExact}) is consistent with
the BBN constraints derived from the primordial ${}^{4}{He}$ and
deuterium abundances, given in Eqs.~(\ref{dhes}) and
(\ref{deuts}), respectively. Therefore, there exists a common
region of the Yukawa parameter space in which both the observed
light-element abundances and the dark matter relic density are
simultaneously reproduced. The overlap of these independent
constraints demonstrates the compatibility of the Yukawa
cosmological model with both BBN and dark matter observations.

Furthermore, the allowed range $-0.017 \lesssim \alpha \lesssim
0.018$ overlaps with constraints reported in the literature from
other astrophysical and cosmological observations
\cite{Benisty2022, Li2014,Deng2009,Borka2013,Jovanovic2023,
Benisty2023}. For comparison, Table~\ref{tab:yukawa_comparison}
summarizes the various observational and analytical constraints on
the Yukawa parameter $\alpha$ (or its analogues $\delta$ and
$\beta$) obtained from these different probes
\cite{Li2014,Deng2009,Borka2013, Benisty2023,
Jovanovic2023,GRAVITY2019,Archidiacono2022,DAgostino2024}.
\begin{table}[H]
    \begin{center}
        \begin{tabular}{|p{4.5cm}|p{4.0cm}|}
            \hline
            \vspace{0.01cm} \textbf{Constraint / Best-Fit} & \textbf{Observational/Analytical Probe} \\ \hline
            $-0.1 < \alpha < 0.12$ & BBN \\ \hline
            $-0.017 < \alpha < 0.018$ & DM freeze-out \\ \hline
            $\alpha\simeq (3.1-5.2)\times10^{-11}$ & Solar System ephemerides \\ \hline
            $\alpha\simeq2.40(2)\times10^{-8}$ & Binary pulsars \\ \hline
            $\delta = 0.017^{+0.0093}_{-0.0053}$ & S-stars (MCMC) \\ \hline
            $0.001 \lesssim \alpha \lesssim 0.05$ & S2 star (scalar field) \\ \hline
            $\alpha\simeq 0.40^{+0.06}_{-0.07}$ & Milky Way rotation \\ \hline
            $\alpha\simeq 0.37^{+0.11}_{-0.17}$ & M31 rotation \\ \hline
            $\alpha\simeq 0.416^{+1.137}_{-0.326}$ (1$\sigma$)& SNe Ia + OHD \\ \hline
            $\alpha \approx 0.04$ & Theoretical Yukawa cosmology \\ \hline
            $\beta < 0.0054$ (95\% C.L.) & Planck CMB + BAO (5F) \\ \hline
            $\beta = 0.0102(28)$ (68\% C.L.) & Planck CMB + $H_0$ (5F) \\ \hline
            $\beta < 0.0287$ (95\% C.L.) & Planck CMB + BAO (CDE) \\ \hline
            $\alpha\simeq 10^{-4}$ to $10^{-5}$  & Solar System planets (Terrestrial planets) \\ \hline
            $\alpha\simeq -5.62\times10^{-3}$  & Solar System planets (Uranus) \\ \hline
            $\alpha = (3.863 \pm 0.373)\times10^{-3}$ & Solar System  \\ \hline
            $\alpha = (4.351 \pm 0.2713)\times10^{-7}$ & Solar System + Cassini  \\ \hline
            $\alpha < 0.581$ & Local Group \\ \hline
        \end{tabular}
    \end{center}
    \caption{Comparison of Yukawa-type coupling constraints from cosmological and astrophysical probes.}
    \label{tab:yukawa_comparison}
    \par\smallskip
    \footnotesize{Note: $\beta$ = dark fifth force strength; $\delta$ = Yukawa interaction strength.}
\end{table}
For the benchmark WIMP considered here, the allowed range of
$\alpha$ is relatively narrow. However, different benchmark
choices for the WIMP mass and annihilation cross section may lead
to a wider allowed region in the $(\alpha,m_\chi,\langle\sigma
v\rangle)$ parameter space. It should be emphasized that  the
numerical bound obtained here depends on the adopted values of the
WIMP mass and annihilation cross section. A more general analysis
would determine the allowed relation among $\alpha$, $m_\chi$ and
$\sigma_0$.
Since the dark matter relic abundance depends on the three
parameters $\alpha$, $m_\chi$ and  $\tilde{\sigma_0}$, fixing the
Yukawa parameter within the BBN-allowed range and imposing the
observational constraint $0.119<\Omega_{\rm cdm}h^2<0.121$
determines the corresponding relation between the WIMP mass and
the annihilation cross section.
\begin{figure}
    \includegraphics[scale=0.84]{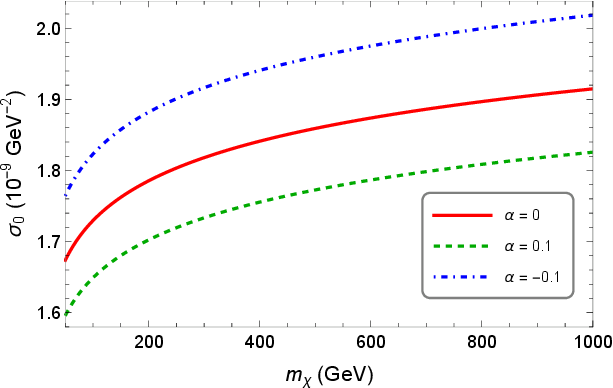}
    \caption{Annihilation cross section $\sigma_0$ vs WIMP mass $m_\chi$ for $\Omega_{\rm cdm}h^2=0.121$ and $\alpha=[-0.1,\;0,\;0.1]$.}
    \label{Fig5}
 \end{figure}
 \begin{figure}
    \includegraphics[scale=0.84]{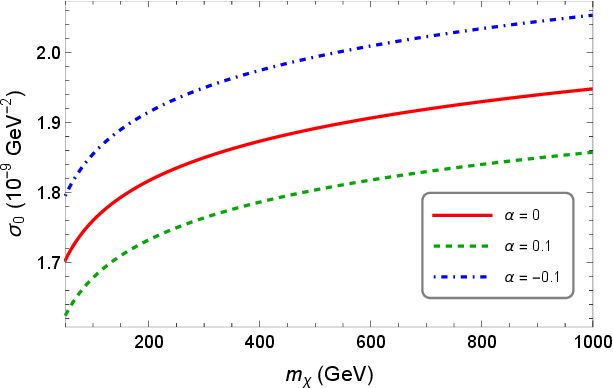}
    \caption{Annihilation cross section $\sigma_0$ vs WIMP mass $m_\chi$ for $\Omega_{\rm cdm}h^2=0.119$ and $\alpha=[-0.1,\;0,\;0.1]$.}
    \label{Fig6}
 \end{figure}
In Fig. \ref{Fig5} and Fig. \ref{Fig6} , we show the annihilation
cross section as a function of the WIMP mass for BBN-allowed
values of $\alpha$ within the range given by
Eq.~(\ref{dhes}). Future CMB experiments, such as CMB-S4 and
the Simons Observatory, are expected to significantly reduce the
uncertainty in $\Omega_{\rm cdm}h^2$, thereby providing more
stringent constraints on $\alpha$ and offering an independent
observational test of the present model.
\section{Time-temperature relation in the early universe through Yukawa cosmology \label{tT}}
\begin{figure}[H]
        \includegraphics[scale=0.9]{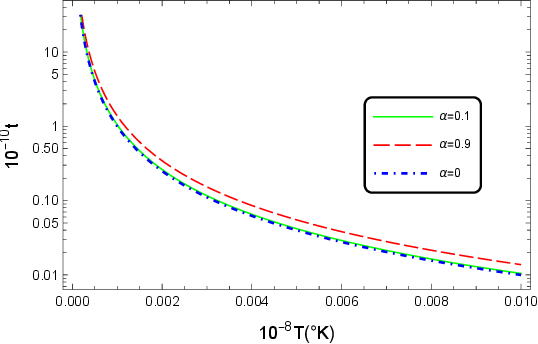}
        \caption{Temperature evolution as a function of cosmic time in the radiation-dominated era of Yukawa cosmology for positive values $\alpha = [0,\; 0.1,\; 0.9 ]$.}
            \label{Fig7}
    \end{figure}

We now turn to the effect of Yukawa cosmology on the relation
between cosmic time and temperature in the early Universe. The
modified Friedmann equations derived earlier alter the
thermodynamic evolution, leading to a time-temperature history
that deviates from the standard cosmological scenario. In the
radiation-dominated epoch, the Universe remains in thermal
equilibrium, which implies entropy conservation within a comoving
volume. This gives rise to the well-known relation \cite{Weinberg}
    \begin{align}\label{ent1}
        s(T)a^3 = \text{const}.
    \end{align}
where $s(T)$ represents the entropy density. The conservation of
entropy per comoving volume imposes a connection between the
thermal history and the expansion of the Universe. Taking the
derivative of Eq.~\eqref{ent1} with respect to cosmic time, we
obtain
    \begin{align}\label{ent2}
        \dot{s}(T)a^3+3\dot{a}a^2s(T)=0 \mapsto
        \frac{ds(T)}{dt}a^3=-3\dot{a}a^2s(T) \;.
    \end{align}
It is convenient to rewrite the above equation in the form
\begin{align}\label{ent2}
        \frac{ds(T)}{dt}=-3Hs(T) \;\; \to \;\; dt=-\frac{ds(T)}{3s(T)}H^{-1} \;.
\end{align}
By plugging the Hubble parameter from Eq.~\eqref{Ffd} into Eq.~\eqref{ent2}, we arrive at
\begin{align}\label{dtT}
        dt=-\frac{ds(T)}{3s(T)}\left( \frac{H_{GR}}{\sqrt{1+\alpha}}\left[1+ \frac{\alpha c^2}{2\lambda^2 H_{GR}^2} \ln\left(\frac{H_{GR}}{1+\alpha}\right)
        \right] \right)^{-1}.
    \end{align}
The integration of the above equation leads to the following expression for cosmic time as a function of temperature
    \begin{eqnarray}\label{dis}
&&  t=-\frac{1}{3}\int_{}^{} \frac{{s}'(T)}{s(T)}\\&&
        \times \left( \frac{H_{GR}}{\sqrt{1+\alpha}} \notag
 \left[1+ \frac{\alpha c^2}{2\lambda^2 H_{GR}^2} \ln\left(\frac{H_{GR}}{1+\alpha}\right)
        \right] \right)^{-1} dT,
    \end{eqnarray}
where a prime denotes differentiation with respect to temperature,
i.e., $f' \equiv df/dT$. In the radiation-dominated epoch,
characterized by the equation of state $p = \rho/3$, the entropy
and energy densities are given by \cite{Weinberg}
\begin{eqnarray}\label{3ss}
    s(T) &=& \frac{2\mathcal{N}a_B T^3}{3},\\
    \rho(T) &=& \frac{\mathcal{N}a_B T^4}{2},\label{ro}
\end{eqnarray}
where $\mathcal{N}$ counts the total relativistic degrees of
freedom, including particles, antiparticles, and their spin states
\cite{Weinberg}. Using the relation $s'(T)/s(T) = 3/T$ together
with some algebraic expansion, Eq.~\eqref{dis} reduces to
    \begin{align}\label{above}
        t=-\int_{}^{}\frac{\sqrt{1+\alpha}}{TH_{GR}} \left[1- \frac{\alpha c^2}{2\lambda^2 H_{GR}^2} \ln\left(\frac{H_{GR}}{1+\alpha}\right)
        \right]dT.
    \end{align}
Inserting the energy density relation from Eq.~\eqref{ro} into Eq.~\eqref{above}, we obtain
    \begin{align}\label{tR}
        t=-\sqrt{1+\alpha}\; \int_{}^{}\left[ \frac{1}{\beta T^3} -  \frac{\alpha c^2}{2\lambda^2 \beta^{3}T^{7}} \ln \left(\frac{\beta T^2}{1+\alpha}\right)\right]dT,
    \end{align}
with the definition $\beta \equiv \sqrt{8\pi G\mathcal{N} a_B / 6c^2}$. Let us first compute the second term on the right-hand side of Eq.~\eqref{tR}
    \begin{align}\label{Ist}
        I= \frac{\alpha c^2}{2\lambda^2 \beta^{3}}\int_{}^{}\frac{ 1}{T^{7}}\ln \left(\frac{\beta T^2}{1+\alpha}\right)dT
        =\frac{\alpha c^2}{2\lambda^2 \beta^{3}} \\
        \times \int_{}^{}\left(\frac{\ln
            \beta}{T^7}+\frac{2\ln T}{T^7}-\frac{\ln(1+\alpha )}{T^7}\right)dT.
    \end{align}
    Applying integration by parts, we find
    \begin{align}\label{Int2}
        I_{}=-\frac{\alpha c^2}{12\lambda^2 \beta^3T^6}\left(\ln \left( \frac{\beta
        T^2}{1+\alpha}\right)+\frac{1}{3}\right)+ \text{const.} \Biggr.
    \end{align}
Combining the above results, we arrive at the following expression for the time-temperature relation in the early Universe within the Yukawa cosmological framework
\begin{multline}\label{tTRe}
    t = \sqrt{1+\alpha} \Biggl[ \frac{1}{2\beta T^2} - \frac{\alpha c^2}{12 \lambda^2 \beta^3 T^6}  \\
\times  \left( \ln \left( \frac{\beta T^2}{1+\alpha} \right) + \frac{1}{3} \right) + \text{const.} \Biggr]
\end{multline}
Equation~\eqref{tTRe} has the correct dimensionality, with the
right-hand side having units of time. Moreover, in the limit
$\alpha = 0$, it reduces to the standard GR expression $t =
\frac{1}{T^2}\sqrt{\frac{3c^2}{16\pi G \mathcal{N} a_B}}$
\cite{Weinberg}. The composition of the early Universe plasma
includes photons, three families of neutrinos and antineutrinos,
and electron-positron pairs, giving $\mathcal{N} = 43/4$ for the
effective number of relativistic degrees of freedom
\cite{Weinberg}. Expressed in cgs units, Eq.~\eqref{tTRe} becomes
    \begin{equation} \label{tTr}
        t=0.994\;\left( {T\over10^{10} K^{\circ}}
        \right)^{-2}F(\alpha,T)+\rm const.,
    \end{equation}
where the correction factor $F(\alpha,T)$ is defined as
    \begin{multline}
        F(\alpha,T)\equiv \left[1- \alpha\;\frac{59535}{T^4}\left[\frac{1}{3}+\ln\left(\frac{5.019\times
        10^{-21}T^2}{1+\alpha}\right)\right] \right] \\
        \times \sqrt{1+\alpha}.
    \end{multline}
As expected, in the limit $\alpha \to 0$, we have $F(\alpha,T) \to
1$, and Eq.~\eqref{tTr} reduces to the standard GR result
\cite{Weinberg}. The influence of the Yukawa modification on the
thermal history of the early Universe is depicted in
Fig.~\ref{Fig7} for  $\alpha > 0$. The figure displays the
temperature $T$ as a function of cosmic time $t$ for different
values of $\alpha$, and clearly illustrates that larger values of
$\alpha$ lead to higher temperatures at a given cosmic time. This
behavior reflects the altered expansion dynamics induced by the
Yukawa correction.
The temperature enhancement observed in Fig.~\ref{Fig7} for
positive values of $\alpha$ can be understood in terms of the
modified gravitational dynamics.In Yukawa cosmology, the Yukawa
modification alters the effective gravitational coupling in the
Friedmann equations, such that $G_{\text{eff}}=G/(1+\alpha)$. For
$\alpha>0$, the effective coupling is reduced  $G_{\text{eff}}<G$,
which corresponds to a slower expansion rate of the early
Universe. As a consequence, the Universe expands more slowly,
allowing its energy density and temperature to decrease more
gradually. Therefore, at a fixed cosmic time, the Universe remains
hotter than in the standard cosmological model. In contrast, for
$-1<\alpha<0$, the effective gravitational coupling becomes larger
than the Newtonian value $G_{\text{eff}}>G$, leading to a faster
expansion rate. Consequently, the energy density and temperature
decline more rapidly, and the Universe becomes cooler than in the
standard cosmological scenario. Hence, the sign of $\alpha$
determines whether the modified gravitational dynamics slows down
or accelerates the cosmic expansion, thereby controlling whether
the early Universe is hotter or cooler relative to the standard
cosmological scenario.
\section{Closing remarks\label{Con}}
BBN is a crucial testing ground for cosmological models. In this
work, we constrain the parameter of Yukawa cosmology using
precision BBN observations and thereby provide a novel test of
modified gravity in the early Universe.

Starting from a Yukawa-type modification to the gravitational
potential, we derive the corresponding corrections to the
Friedmann equations through the first law of thermodynamics on the
apparent horizon and obtain a modified Friedmann equation with a
logarithmic correction term. The modified expansion history
depends on two parameters: the coupling constant $\alpha$ and the
wavelength $\lambda$ (or equivalently the graviton mass $m_g =
\hbar/(\lambda c)$). Using the latest observational data on
primordial light-element abundances, we have established stringent
constraints on $\alpha$.

Our analysis reveals that ${}^{4}{He}$ measurements constrain
$\alpha$ to $-0.24 \lesssim \alpha \lesssim 0.12$, while deuterium
allows a wider range of $-0.5 \lesssim \alpha \lesssim 1.5$. The
consistency between these two independent constraints indicates
that the Yukawa cosmological model remains viable for sufficiently
small $|\alpha|$, as it is able to reproduce both the ${}^{4}{He}$
and deuterium abundances. For ${}^{7}{Li}$, we obtain $-0.76
\lesssim \alpha \lesssim -0.72$, which shows no overlap with the
${}^{4}{He}$ and deuterium ranges. This reflects the well-known
cosmological \textit{Lithium Problem}, which persists in this
framework as no single value of $\alpha$ can simultaneously
satisfy all three elemental abundances.

Extending our analysis to the dark matter sector, we investigated
the thermal freeze-out of WIMPs within the Yukawa cosmology
framework. Solving the modified Boltzmann equation that
incorporates the Yukawa correction to the Hubble parameter, we
derived an analytical expression for the relic abundance as a
function of $\alpha$. Requiring consistency with the observed dark
matter relic density $\Omega_{\rm CDM}h^2 = 0.120 \pm 0.001$ for a
benchmark thermal WIMP yields the independent constraint $-0.017
\lesssim \alpha \lesssim 0.018$. This constraint lies entirely
within the BBN-allowed range, which demonstrates consistency
between these independent cosmological probes. The final combined
constraint, dominated by the more restrictive WIMP bound, is
therefore $-0.017 \lesssim \alpha \lesssim 0.018$.

Furthermore, we have shown that for fixed values of $\alpha$
within the BBN-allowed range, the observational constraint on the
dark matter abundance determines a corresponding relation between
the WIMP mass $m_\chi$ and the annihilation cross section
$\sigma_0$ (Fig.~\eqref{Fig5} and Fig.~\eqref{Fig6}). This
provides a predictive framework for testing Yukawa cosmology
against future dark matter searches, including direct detection
experiments.

Additionally, we have derived the cosmic time-temperature relation
in Yukawa cosmology. For $\alpha > 0$, the modified expansion rate
reduces the effective Newton constant ($G_{\text{eff}} =
G/(1+\alpha)$), which slows the expansion and allows the Universe
to remain hotter for longer durations at a given cosmic time.
Conversely, for $\alpha < 0$, the expansion is accelerated,
leading to a cooler Universe at the same cosmic time.

In summary, our results demonstrate that the early Universe,
through BBN and dark matter relic abundance, imposes powerful and
independent constraints on Yukawa cosmology. The consistency
between these independent probes reinforces the viability of the
framework. These constraints complement and extend bounds obtained
from late-time cosmological probes and local laboratory
experiments, and thus offer a more complete picture of the
viability of Yukawa-type modifications to gravity.
\acknowledgments{We are grateful to Shiraz University Research
Council.}


\begin{thebibliography}{99}
\bibitem{Sakr2022} Z. Sakr, S. Ilić and A. Blanchard, \textit{Cluster counts - III. $\Lambda$CDM extensions and the cluster tension}, Astron. Astrophys {\bf666}, A34 (2022), [arXiv:2112.14171].
\bibitem{Akarsu2023} \"O. Akarsu, E. Di Valentino, S. Kumar, R. C. Nunes, J. A. Vazquez and A. Yadav, \textit{$\Lambda_{\rm s}$CDM model: A promising scenario for alleviation of cosmological tensions}, (2023), [arXiv:2307.10899].
\bibitem{Turner2026} M. S. Turner, \textit{Everyone wants something better than $\Lambda$CDM}, Proc. Natl. Acad. Sci. U.S.A. {\bf123}, 8 (2026), [arXiv:2510.05483].
\bibitem{Blanchard2025} A. Blanchard, \textit{The fallacies of LCDM falsifications}, in \textit{Cosmology Research - Addressing Current Problems with Astrophysics}, IntechOpen \textbf{135} (2025), [arXiv:2505.06244].
\bibitem{Koyama2016} K. Koyama, \textit{Cosmological tests of modified gravity}, Rep. Prog. Phys {\bf79}, 4 (2016), [arXiv:1504.04623].
\bibitem{Clifton2012} T. Clifton, P. G. Ferreira, A. Padilla and C. Skordis, \textit{Modified gravity and cosmology}, Phys. Rept {\bf513}, 1 (2012), [arXiv:1106.2476].
\bibitem{Carroll2005} S. M. Carroll, A. De Felice, V. Duvvuri, D. A. Easson, M. Trodden and M. S. Turner, \textit{Cosmology of generalized modified gravity models}, Phys. Rev. D {\bf71}, 063513 (2005), [arXiv:astro-ph/0410031].
\bibitem{Nishonov2026} I. Nishonov, B. Rahmatov, S. U. Khan, M. Zahid, J. Rayimbaev, I. Ibragimov and E. Davletov, \textit{Yukawa black holes in modified gravity: From thermodynamics to particle collisions}, Ann. Phys {\bf489}, 170332 (2026).
\bibitem{DeMartino2018} I. De Martino, R. Lazkoz and M. De Laurentis, \textit{Analysis of the Yukawa gravitational potential in f(R) gravity. I. Semiclassical periastron advance}, Phys. Rev. D {\bf97}, 104067 (2018), [arXiv:1801.08135].
\bibitem{AraujoFilho2024} A. A. Araújo Filho, K. Jusufi, B. Cuadros-Melgar, G. Leon, A. Jawad and C. E. Pellicer, \textit{Charged black holes with Yukawa potential}, Phys. Dark. Univ {\bf46}, 101711 (2024), [arXiv:2401.15211].
\bibitem{DelValle2018} J. C. del Valle and D. J. Nader, \textit{Towards the theory of the Yukawa potential}, J. Math. Phys {\bf59}, 102103 (2018), [arXiv:1807.11898].
\bibitem{Ver}  E. Verlinde, \textit{On the Origin of Gravity and the Laws of Newton}, JHEP {\bf1104}, 029 (2011), [arXiv:1001.0785].
\bibitem{Cai2} R. G. Cai, L. M. Cao and N. Ohta,
\textit{Friedmann equations from entropic force,} Phys. Rev. D
{\bf81}, 061501 (2010), [arXiv:1001.3470].
\bibitem{sheyECFE} A. Sheykhi, \textit{Entropic corrections to Friedmann equations}, Phys. Rev. D {\bf81}, 104011 (2010), [arXiv:1004.0627].
\bibitem {KS} K. Jusufi, A. Sheykhi, \textit{Entropic corrections to Friedmann equations and bouncing universe due to the zero-point length}, Phys.  Lett. B \textbf{836}, 137621 (2023), [arXiv:2210.01584].
\bibitem{Sheykhi:2025} A. Sheykhi, L. Liravi and K. Jusufi, \textit{Thermodynamical properties of nonsingular universe}, Phys. Dark. Univ {\bf48}, 101931 (2025), [arXiv:2407.21426].
\bibitem{Cai22} M.~Akbar and R.~G.~Cai, \textit{Thermodynamic behavior of the Friedmann equation at the apparent horizon of the FRW universe},
Phys. Rev. D {\bf 75}, 084003 (2007), [arXiv:hep-th/0609128].
\bibitem{wang1} B. Wang, E. Abdalla and R. K. Su, \textit{Relating Friedmann equation to Cardy formula in universes with cosmological constant},
Phys. Lett. B {\bf503},  394 (2001), [arXiv:hep-th/0101073].
\bibitem{RH1} R. H. Cyburt, B. D. Fields and K. A. Olive, \textit{Primordial Nucleosynthesis with CMB Inputs: Probing the Early Universe and Light Element Astrophysics},
Astropart. Phys {\bf17}, 87 (2002), [arXiv:astroph/0105397].

\bibitem{RH2} Richard H. Cyburt, et. al., \textit{Big-Bang Nucleosynthesis:2015}, Rev. Mod. Phys {\bf88}, 015004 (2016), [arXiv:1505.01076].

\bibitem{Planck}  Planck Collaboration,  \textit{Planck 2018 results. VI. Cosmological parameters}, Astron. Astrophys {\bf641}, A 6 (2020), [arXiv:1807.06209].

\bibitem {Brain} B. D. Fields, K. A. Olive, T.H. Yeh, C. Yung, \textit {Big-Bang nucleosynthesis after Planck}, J. Cosm. Asto. Phys \textbf{03}, 010 (2020), [arXiv:1912.01132].
\bibitem{Anish} A. Ghoshal, G. Lambiase, \textit{Constraints on Tsallis cosmology from big-bang nucleosynthesis and dark matter freeze-out}, [arXiv:2104.11296].

\bibitem{Luciv} G. G. Luciano, \textit {Modified Friedmann equations from Kaniadakis entropy and cosmological implications on baryogenesis and ${_{}^{7}\textrm{Li}}$-abundance},
Eur. Phys. J. C \textbf{82}, 314 (2022).

\bibitem{Sahoo2} S. S. Mishra, A. Kolhatkar, P.K. Sahoo, \textit{Big-Bang Nucleosynthesis constraints on $f(T,T)$ Gravity}, Phys. Lett. B {\bf848}, 138391 (2024), [arXiv:2312.07558].
\bibitem{Ava1} A. Sheykhi and A. Shabazi Sooraki, \textit{Barrow cosmology and big-bang nucleosynthesis}, Phys. Rev. D {\bf111}, 043518 (2025), [arXiv:2411.06075].
\bibitem{Coc} Coc et. al, \textit{Standard Big-Bang Nucleosynthesis up to CNO with an improved extended nuclear network}, Astro. Phys. J {\bf744}, 158 (2012),
[arXiv:1107.1117].
\bibitem{Bertone:2004pz}G.~Bertone, D.~Hooper and J.~Silk,\textit{Particle dark matter: Evidence, candidates and constraints}, Phys.\ Rept\  {\bf 405}, 279 (2005), [arXiv: hep-ph/0404175].
\bibitem{Feng:2010gw}
J.~L.~Feng,
\textit{Dark Matter Candidates from Particle Physics and Methods of Detection},
Ann.\ Rev.\ Astron.\ Astrophys\  {\bf 48}, 495 (2010),
[arXiv:1003.0904].

\bibitem{Gelmini:2015zpa}
G.~Gelmini,
\textit{The hunt for dark matter,}
in Theor.\ Adv.\ Stud.\ Inst.\ Elem.\ Part.\ Phys: Journeys Through the Precision Frontier: Amplitudes for Colliders,
WSPC \textbf{559} (2015), [arXiv:1502.01320].
\bibitem{Roszkowski:2017nbc}
L.~Roszkowski, E.~M.~Sessolo and S.~Trojanowski,
\textit{WIMP dark matter candidates and searches—current status and future prospects},
Rept.\ Prog.\ Phys\  {\bf 81}, 066201 (2018),
[arXiv:1707.06277].
\bibitem{Arcadi:2017kky}
G.~Arcadi, M.~Dutra, P.~Ghosh, M.~Lindner, Y.~Mambrini, M.~Pierre, S.~Profumo and F.~S.~Queiroz,
\textit {The waning of the WIMP? A review of models, searches, and constraints},
Eur.\ Phys.\ J.\ C {\bf 78}, 203 (2018)
[arXiv:1703.07364].
\bibitem{Battaglieri:2017aum}
M. Battaglieri. et. al,
\textit{US Cosmic Visions: New Ideas in Dark Matter 2017: Community Report}, FERMILAB. CONF \textbf{17}, 282
[arXiv:1707.04591].

\bibitem {Munoz:2003gx}C.~Munoz, \textit{Dark matter detection in the light of recent experimental results}, Int. J. Mod. Phys. A \textbf{19}, 3093 (2004),
[arXiv:hep-ph/0309346Xi].

\bibitem {Taoso:2007qk}
M.~Taoso, G.~Bertone and A.~Masiero,
\textit{Dark matter candidates: A ten-point test},
JCAP \textbf{03}, 022 (2008),
[arXiv:0711.4996 [astro-ph]].

\bibitem {Jungman:1995df}
G.~Jungman, M.~Kamionkowski and K.~Griest,
\textit{Supersymmetric dark matter},
Phys. Rept \textbf{267}, 195 (1996),
[arXiv:hep-ph/9506380].
\bibitem {Mod90} M. Drees, H. Iminniyaz and M. Kakizaki, \textit{Constraints on the very early universe from thermal WIMP dark matter}, Phys. Rev. D {\bf76}, 103524 (2007), [arXiv:0704.1590].
    \bibitem {Kolb:1988aj}
E.~W.~Kolb and M.~S.~Turner,
\textit{The Early Universe},
Frontiers in Physics \textbf{70}, Addison-Wesley (1990).
\bibitem{Catena:2004ba}
R. Catena, N. Fornengo, A. Masiero, M. Pietroni and F. Rosati,
\textit{Dark Matter Relic Abundance and Scalar-Tensor Dark Energy},
Phys.\ Rev.\ D {\bf 70}, 063519 (2004), [arXiv:astro-ph/0403614].

\bibitem{Catena:2006bd}
R. Catena, M. Pietroni and L. Scarabello,
\textit{Einstein and Jordan frames reconciled: a frame-invariant approach to scalar-tensor cosmology}, Phys.\ Rev.\ D {\bf 76}, 8 (2007), [arXiv:astro-ph/0604492].

\bibitem{Kang} J. U. Kang and G. Panotopoulos, \textit{Big-bang nucleosynthesis and WIMP dark matter in modified gravity}, Phys. Lett. B \textbf{677}, 6 (2009), [arXiv:0811.2885].

\bibitem{Salati:2002md}
P. Salati,
\textit{Quintessence and the Relic Density of Neutralinos},
Phys.\ Lett.\ B {\bf 571}, 131 (2003),
[arXiv:astro-ph/0207396].

\bibitem{Profumo:2003hq}
S. Profumo and P. Ullio,
\textit{SUSY Dark Matter and Quintessence},
JCAP {\bf 11}, 006 (2003), [arXiv:hep-ph/0309220].

\bibitem{Okada:2004nc}
N. Okada and O. Seto,
\textit{Relic density of dark matter in brane world cosmology},
Phys.\ Rev.\ D {\bf 70}, 8 (2004), [arXiv:hep-ph/0407092].

\bibitem{Bernal:2018ins}
N. Bernal, C. Cosme and T. Tenkanen,
\textit{Phenomenology of Self-Interacting Dark Matter in a Matter-Dominated Universe},
Eur.\ Phys.\ J.\ C {\bf 79}, 99 (2019),
[arXiv:1803.08064 [hep-ph]].
\bibitem{Jusufi2023} K. Jusufi, G. Leon and A. D. Millano, \textit{Dark Universe phenomenology from Yukawa potential?}, Phys. Dark. Univ. {\bf42}, 101318 (2023), [arXiv:2304.11492].
    \bibitem{Sheykhi:2021}
A. Sheykhi, \textit{Barrow entropy corrections to Friedmann equations}, Phys. Rev. D {\bf103}, 123503 (2021), [arXiv:2102.06550].
    \bibitem{Padmanabhan:2004}
T. Padmanabhan, \textit{Entropy of static spacetimes and microscopic density of states}, Class. Quantum. Grav {\bf21}, 4485 (2004), [arXiv:gr-qc/0308070].

\bibitem{Visser:1998}
M. Visser, \textit{Mass for the graviton}, Gen. Rel. Grav {\bf30}, 12 (1998), [arXiv:gr-qc/9705051].

\bibitem{Capozziello:2009}
S. Capozziello, A. Stabile and A. Troisi, \textit{A general solution in the Newtonian limit of f(R)-gravity}, Mod. Phys. Lett. A {\bf24}, 659 (2009), [arXiv:0901.0448]
\bibitem{Berezhiani2009} Z. Berezhiani, F. Nesti, L. Pilo and N. Rossi, \textit{Gravity modification with Yukawa-type potential: dark matter and mirror gravity}, JHEP {\bf07}, 083 (2009), [arXiv:0902.0144].

\bibitem{Benisty2022} D. Benisty, \textit{Testing modified gravity via Yukawa potential in two body problem: Analytical solution and observational constraints}, Phys. Rev. D {\bf106}, 043001 (2022), [arXiv:2207.08235].
    \bibitem{Chatterjee:2020}
A. Chatterjee and A. Ghosh, \textit{Exponential corrections to black hole entropy}, Phys. Rev. Lett. {\bf125}, 041302 (2020), [arXiv:2007.15401].
\bibitem{Hay1}
S. A. Hayward, \textit{Unified first law of black-hole dynamics and relativistic thermodynamics}, Class. Quant. Grav \textbf{15}, 3147 (1998), [arXiv:gr-qc/9710089].

\bibitem{Hay2}
S. A. Hayward, S. Mukohyama, and M. C. Ashworth \textit{Ashworth, Dynamic black-hole entropy}, Phys. Lett. A \textbf{256}, 347 (1999), [arXiv:gr-qc/9810006].

\bibitem{Bak}
D. Bak and S. J. Rey, \textit{Cosmic Holography}, Class. Quant. Grav \textbf{17}, L83 (2000), [arXiv:hep-th/9902173].






\bibitem {Luciano}  G. G. Luciano, \textit {Primordial big-bang nucleosynthesis and generalized uncertainty principle} Eur. Phys. J. C \textbf{81}, 1086 (2021), [arXiv:2111.06000].
\bibitem {Boran}  S. Boran and E. O. Kahya,\textit {Testing a dilaton gravity model using nucleosynthesis}, Adv. High. Energy. Phys \textbf {1}, 282675 (2014), [arXiv:1310.6145].
\bibitem {Adv} G. Steigman, \textit {Neutrinos and big-bang nucleosynthesis}, Adv. High. Energy. Phys \textbf {1}, 268321 (2012), [arXiv:1208.0032].

\bibitem {Simha} V. Simha, G. Steigman, \textit {Constraining the early-Universe baryon density and expansion rate}, J. Cosm. Astro. Phys \textbf {06}, 016 (2008), [arXiv:0803.3465].

\bibitem {Sahoo} S. Bhattacharjee, P. K. Sahoo, \textit {Big-bang nucleosynthesis and entropy evolution in $f (R, T)$ gravity}, Eur. Phys. J. Plus \textbf{135}, 350 (2020), [arXiv:2004.04684].
\bibitem {Kneller} J. P. Kneller, G. Steigman, \textit {BBN for pedestrians}, NJP \textbf{6}, 117 (2004), [arXiv:astro-ph/0406320].

\bibitem {Annu} G. Steigman, \textit {Primordial nucleosynthesis in the precision cosmology era}, Annu.\ Rev.\ Nucl.\ Part.\ Sci  \textbf{57}, 463 (2007), [arXiv:0712.1100].

\bibitem {Wamp} N. Jarosik, C. L. Bennett, J. Dunkley, B. Gold, M. R. Greason, M. Halpern, R. S. Hill. et. al, \textit{Seven-year wilkinson microwave anisotropy probe
    (WMAP*) observations: Sky maps, systematic errors, and basic results}, Astrophys. J. Suppl \textbf{192}, 18 (2011), [arXiv:1001.4758].

    \bibitem{Gonzalez2023} E. Gonzalez, K. Jusufi, G. Leon and E. N. Saridakis, \textit{Observational constraints on Yukawa cosmology and connection with black hole shadows}, Phys. Dark. Univ {\bf42}, 101304 (2023), [arXiv:2305.14305].

    \bibitem{AboHasan2023} N. Abo Hasan, N. Joudieh and N. Chamoun, \textit{Dynamics and stability of the two-body problem with Yukawa correction to Newton's gravity, revisited and applied numerically to the solar system}, Universe {\bf9}, 45 (2023) [arXiv:2301.02498v1].

    \bibitem {Feng:2003zu}
J. L. Feng, \textit{Supersymmetry and cosmology}, Ann. Phys {\bf315}, 2 (2005), [arXiv:hep-ph/0405215].


    \bibitem{Dodelson} S. Dodelson and F. Schmidt, \textit{Modern Cosmology}, Elsevier (2024).
\bibitem{Husdal} L. Husdal, \textit{On effective degrees of freedom in the early universe}, Galaxies \textbf{4}, 78 (2016), [arXiv:1609.04979].
\bibitem{Laine} M. Laine and Y. Schr\"oder, \textit{Quark mass thresholds in QCD thermodynamics}, Phys. Rev. D \textbf{73}, 085009 (2006), [arXiv:hep-ph/0603048].
\bibitem{Profumo}
S. Profumo, L. Giani, and O. F. Piattella,
\textit{An Introduction to Particle Dark Matter},
Universe \textbf{5}, 213 (2019),
[arXiv:1910.05610].
\bibitem{Smirnov}
J. Smirnov and J. F. Beacom,
\textit{TeV-Scale Thermal WIMPs: Unitarity and its Consequences},
Phys. Rev. D \textbf{100}, 043529 (2019),
[arXiv:1904.11503].

\bibitem{Steigman}
G. Steigman, B. Dasgupta, and J. F. Beacom,
\textit{Precise Relic WIMP Abundance and its Impact on Searches for Dark Matter Annihilation},
Phys. Rev. D \textbf{86}, 023506 (2012),
[arXiv:1204.3622].

\bibitem{Drees}M. Drees, F. Hajkarim, and E. R. Schmitz, \textit{The Effects of QCD Equation of State on the Relic Density of WIMP Dark Matter}, Phys. Rev. D \textbf{92}, 063525 (2015),[arXiv:1503.03513].
\bibitem{Leane}R. K. Leane, T. R. Slatyer, J. F. Beacom, and K. C. Y. Ng, \textit{GeV-Scale Thermal WIMPs: Not Even Slightly Ruled Out}, Phys. Rev. D \textbf{98}, 023016 (2018),
[arXiv:1805.10305].

\bibitem{Li2014} Z. W. Li, S. F. Yuan, C. Lu and Y. Xie, \textit{New upper limits on deviation from the inverse-square law of gravity in the solar system: a Yukawa parameterization}, Res. Astron. Astrophys {\bf14}, 139 (2014).

\bibitem{Deng2009} X. M. Deng, Y. Xie and T. Y. Huang, \textit{Modified scalar-tensor-vector gravity theory and the constraint on its parameters}, Phys. Rev. D {\bf79}, 044014 (2009). [arXiv:0901.3730v1].

\bibitem{Borka2013} D. Borka, P. Jovanović, V. Borka Jovanović and A. F. Zakharov, \textit{Constraining the range of Yukawa gravity interaction from S2 star orbits}, JCAP {\bf11}, 050 (2013), [arXiv:1311.1404].

\bibitem{Jovanovic2023} P. Jovanović, V. Borka Jovanović, D. Borka and A. F. Zakharov, \textit{Constraints on Yukawa gravity parameters from observations of bright stars}, JCAP {\bf03}, 056 (2023), [arXiv:2211.12951].

\bibitem{Benisty2023} D. Benisty and S. Capozziello, \textit{Tracking the Local Group dynamics by extended gravity}, Phys. Dark. Univ {\bf39}, 101175 (2023), [arXiv:2301.06614].
\bibitem{GRAVITY2019} GRAVITY Collaboration, A. Amorim et al., \textit{Scalar field effects on the orbit of S2 star}, Mon. Not. R. Astron. Soc. {\bf489}, 4606 (2019), [arXiv:1908.06681].

\bibitem{Archidiacono2022} M. Archidiacono, E. Castorina, D. Redigolo and E. Salvioni, \textit{Unveiling dark fifth forces with linear cosmology}, JCAP {\bf10}, 074 (2022), [arXiv:2204.08484].

\bibitem{DAgostino2024} R. D'Agostino, K. Jusufi and S. Capozziello, \textit{Testing Yukawa cosmology at the Milky Way and M31 galactic scales}, Eur. Phys. J. C {\bf84}, 386 (2024), [arXiv:2404.01846].

\bibitem{Weinberg} S. Weinberg, \textit {Cosmology, New York: Oxford University press}, (2008).


\end{thebibliography}
\end{document}